\documentclass[namedreferences]{solarphysics}
\usepackage[optionalrh]{spr-sola-addons}
\usepackage{threeparttable}
\usepackage{graphicx}
\usepackage{graphics}
\usepackage{color}           
\usepackage{url}             

\newcommand{\arcsec}{$^{\prime \prime}$~}
\newcommand{\farcs}{$.^{\prime \prime}$}

\begin{document}
\begin{article}
\begin{opening}

\title{Comparative Analysis of a Transition Region Bright Point with a Blinker and Coronal Bright Point Using Multiple EIS Emission Lines}

%
\author{N. Brice Orange, Hakeem M. Oluseyi, David L. Chesny, Maulik Patel, Katie Hesterly, Lauren Preuss, Chantale Neira, and Niescja E. Turner}
\runningtitle{Transition Region Bright Point}
\runningauthor{Orange et al.}
\institute{Department of Physics \& Space Sciences, Florida Institute of Technology, Melbourne, FL  32901}

\begin{abstract}

Since their discovery twenty year ago, transition region bright points have never been observed spectroscopically. Bright point properties have not been compared with similar transition region and coronal structures. In this work we have investigated three transient quiet Sun brightenings including a transition region bright point (TR BP), a coronal bright point (CBP) and a blinker. We use time-series observations of the extreme ultraviolet emission lines of a wide range of temperature $T$ ($\log T = 5.3 - 6.4$) from the {\it EUV imaging spectrometer} (EIS) onboard the {\it Hinode} satellite. We present the EIS temperature maps and Doppler maps, which are compared with magnetograms from the Michelson Doppler Imager (MDI) onboard the SOHO satellite. Doppler velocities of the TR BP and blinker are $\le$\,25 km s$^{-1}$,
which is typical of transient TR phenomena. The Dopper velocities of the CBP were found to be $\le$\,20 km s$^{-1}$ with exception of those measured at $\log T$\,$=$ 6.2 where a distinct bi-directional jet is observed. From an EM loci analysis we find evidence of single and double isothermal components in the TR BP and CBP, respectively. TR BP and CBP loci curves are characterized by broad distributions suggesting the existence of unresolved structure. By comparing and contrasting the physical characteristics of the events we find the BP phenomena are an indication of multi-scaled self similarity, given similarities in both their underlying magnetic field configuration and evolution in relation to EUV flux changes. In contrast, the blinker phenomena and the TR BP are sufficiently dissimilar in their observed properties as to constitute different event classes. Our work is an indication that the measurement of similar characteristics across multiple event types holds class-predictive power, and is a significant step towards automated solar atmospheric multi-class classification of unresolved transient EUV sources. Finally, the analysis performed here establishes a connection between solar quiet region CBPs and jets.
\end{abstract}
\end{opening}
%

\section{Introduction}
\label{sec:intro}
Compact stochastic brightenings have been observed throughout the solar atmosphere. Events known as bright points (BP) appear to be ubiquitous across all temperature regimes of the solar atmosphere \cite{1996ApJ...466..529K,1997ApJ...491..952K}. First discovered in early rocket flights using grazing-incidence optics \cite{Krieger1971}, BPs are among the most widespread and abundant forms of solar activity. They have been observed to occur in active regions ({\it e.g.}, \opencite{Romanoetal2012SoPh}), coronal holes \cite{Habbaletal1990ApJ,Subramanianetal2010A&A,Doscheketal2010ApJ}, regions of quiet Sun \cite{Habbaletal1990ApJ,Abramenkoetal2010ApJ,Sanchezetal2010ApJ}, and are usually associated with the magnetic network \cite{Golub1974,EgamberdievIakovkin1983DoUzb,1997ApJ...491..952K}. In all cases BPs occur in association with mixed magnetic flux that undergoes cancellation or rearrangement while being dynamically driven by hydrodynamic flows or flux emergence (P.M. Harvey {\it et al.} 1985; \opencite{Webbetal1993SoPh}; K.L.Harvey {\it et al.} 1994; \opencite{1994ApJ...427..459P}; \opencite{1997ApJ...491..952K}), suggesting that the BP phenomena is a consequence of magnetic reconnection. Indeed, BPs have often been described as being tiny solar active regions \cite{2001ApJ...553..429L}.

The  surface densities of BPs appears to increase with decreasing temperature \cite{2001SoPh..198..347Z,2005SoPh..228..285M}, although the exact functional form remains unknown. Nonetheless, even near the highest temperatures where BPs are observed --- and hence away from the peak of their population distribution --- BPs have been measured to contribute significantly to the solar irradiance. For example, a survey of X-ray BPs (XBPs) using data from {\it Yohkoh}/SXT, concluded that these hot XBPs covered $\approx$\,1.4\% of the quiet-Sun surface area and contributed $\approx$\,5\% of the quiet-Sun X-ray irradiance \cite{2001SoPh..198..347Z}. Naive extrapolation of the BP population distribution to lower temperatures suggests that BPs should be even more important contributors to solar irradiance at lower temperatures, such as in the upper TR ($5.0 \leq \log~T < 6.0$).

The first studies of transition region (TR) BPs were presented by Habbal, Dowdy, and Withbroe (1990) and \citeauthor{1993ApJ...411..406D} (\citeyear{1993ApJ...411..406D}), who compared lower TR BPs observed in the emission of C\,\textsc{iii}~$\lambda$977.02 ($\log T = 4.8$) to coronal BPs observed in the emission of Mg\,\textsc{x}~$\lambda$624.94 ($\log T = 6.8$). Dowdy (1993) utilized spectroheliograms in the emission of O\,\textsc{vi}~$\lambda$1031.91 ($\log T = 5.4$), Ne\,\textsc{vii}~$\lambda$465.22 ($\log T = 5.7$), and Mg\,\textsc{x}~$\lambda$624.94 ($\log T = 6.8$) to identify BPs at upper TR temperatures. The focus of these early studies was to determine whether the BPs studied existed thermally and magnetically isolated from the large-scale corona. Both studies confirmed this question.

In an effort to better constrain the distribution of BPs across the upper TR, \citeauthor{Bruni2006} (\citeyear{Bruni2006}) performed a measurement of TR BP surface number densities and radiated powers at two TR temperatures and in three heliographic latitude bands. He used large area upper TR spectroheliograms in the emission of N\,\textsc{v}~$\lambda$1238.81 ($\log T = 5.2$) and Ne\,\textsc{viii}~$\lambda$770.41 ($\log T = 5.8$) obtained by the {\it Solar Ultraviolet Measurements of Emitted Radiation} (SUMER; \opencite{Wilhelmetal1995SoPh}) spectrometer on 7--8 June 1996 near solar minimum. The flux-calibrated SUMER data were searched for compact brightenings above magnetic bipoles. Bruni (2006) found that the BP contribution was $\approx$\,$20\%$ in the quiet Sun and $\approx$\,$15\%$ in a polar coronal hole at $\log T = 5.2$ and $\approx$\,$10\%$ in the quiet-Sun and $\approx$\,$5\%$ in the polar region at $\log T = 5.8$. The surface densities of TR BPs observed in N\,\textsc{v} was $\approx$\,$1$\,$\times$\,$10^{-3}$\,Mm$^{-2}$ and in Ne\,\textsc{viii} it was $\approx$\,$5$\,$\times$\,$10^{-4}$\,Mm$^{-2}$. These surface densities may be compared with the surface densities of  coronal BPs observed in the SOHO {\it Extreme-Ultraviolet Imaging Telescope} (EIT; \opencite{Delaboudiniereetal1995SoPh}) 171\AA~filtergram by \citeauthor{2001ApJ...553..429L}~(\citeyear{2001ApJ...553..429L}), which was $1.3$\,$\times$\,$10^{-4}$\,Mm$^{-2}$. If confirmed, Bruni's measurements of surface densities and irradiance contributions are consistent with the trend in the BP distribution seen at coronal temperatures. Bruni (2006) noted that his surface density counting errors due to Type I (false-negative) and Type II (false-positive) errors was $\approx$\,30\%. The Type II errors were most likely due to contamination by compact, TR transients known as blinkers
\cite{1997SoPh..175..467H,1999A&A...351.1115H,2001A&A...373.1056B,2002SoPh..206...21B,2002SoPh..206..249P,2004A&A...418L...9D}.

Just as coronal BPs have been associated with magnetic flux cancellation associated with both converging magnetic bipoles \cite{1976SoPh...50..311G,Webbetal1993SoPh,1994ApJ...427..459P} and emerging magnetic flux \cite{Strongetal1992PASJ}, upper TR BPs and blinkers may form a similar dichotomy --- depending on how one defines a blinker. Recent studies have suggested that the blinker classification is generally plagued by Type II errors and that a range of phenomena have been grouped together under the blinker designation \cite{2003A&A...409..755H}. In this paper we are thus restricting the blinker designation to the types of phenomena observed by Doyle, Roussev, and Madjarska (2004) and others, which are density enhancement events generated by reconnection involving cool plasma tied to emerging magnetic flux and hot coronal plasma confined in pre-existing coronal structures. Likewise, we will define TR BPs as being upper TR structures generated in association with colliding and cancelling photospheric magnetic flux elements.

There is much to be learned from a study combining TR BPs, coronal BPs, and blinkers. First, individual TR BPs have never been observed in multiple emission lines nor have their Doppler signatures been measured. The physics of TR BPs may very well be similar to that of coronal BPs as an additional manifestation of multi-scaled self-similarity in the solar atmosphere \cite{Vlahosetal2004ApJ,Raouafietal2010ApJ}. Coronal BPs display a wealth of enigmatic dynamic behaviors that if found in TR BPs might provide new insights into plasma heating and acceleration. For example, coronal BPs show periodic flashes or flaring for which there is not currently a consensus model \cite{Zhangetal2012ApJ,2011A&A...529A..21K}. The larger of the flaring BPs often exhibit eruptive behavior in the form of jets, thus contributing mass to the solar wind \cite{Madjarskaetal2012A&A}. It has been inferred that some fundamental plasma processes ({\it e.g.}, the efficiency of magnetic reconnection) varies across the TR \cite{1998ApJ...507..433L,LongcopeKankelborg1999ApJ,Bruni2006}. Therefore, observing a similar physical process under different plasma conditions may be insightful. On the other hand, blinkers, which are energized by the reconnection of emerging and pre-existing flux, and TR BPs, which are energized by reconnection driven by hydrodynamic flows, both occur under the same TR conditions. Comparing these different processes occurring under similar plasma parameters may also yield new insights. Finally, solar physics is rapidly moving to a new operational mode where machine learning based software is replacing humans as data processors, a trend that will only increase into the future. These software tools require a set of defining event ``features" to distinguish classes. Blinkers have already proven to be a classification challenge for human operators \cite{2003A&A...409..755H,Subramanianetal2012A&A}. Obtaining robust probabilistic classifications of compact transients in general will be a major challenge in the coming years. A comparative study may identify distinguishing features that may be useful in this effort.

The purpose of this paper is to perform the first multi-wavelength spectroscopic study of an individual TR BP and also to perform the first comparative study of TR BPs with coronal BPs and blinkers. The next section presents an overview of the observational data used in our study and its processing. Section~\ref{sec:measurements} describes our techniques for identifying the events we study and measuring their morphologies, line fluxes, Doppler velocity maps, and temperature structures. Section~\ref{sec:results} presents the measurement results and their implications for each event. The discussion of our results and a brief summary are provided in Section~\ref{sec:discussion}.

\section{Observations}
\label{sec:observations}
To investigate the spectroscopic properties of the BPs we study, data were obtained from the {\it EUV Imaging Spectrometer} (EIS; \opencite{Culhaneetal2007SoPh}) on the \textit{Hinode} satellite. The EIS data was taken from 11:13:11 UT to 23:12:40 UT at $\approx$ 45 min intervals on 22 January 2008. The data consists of raster scans with a 2\arcsec slit width using 1\arcsec steps resulting in a final field-of-view (FOV) of 119\farcs8\,$\times$\,360\arcsec. A total of ten emission lines with coverage from the upper TR ($\log T = 5.4$) to the corona ($\log T = 6.2$) were used. The emitting ions, the wavelengths of the emission lines, the atomic transitions of the emission lines, and their peak formation temperatures are all given in Table~\ref{tbl:eis_elines}.

The EIS level-0 data were processed and calibrated via the standard \textrm{Solar SoftWare} (SSW) routine {\sf  eis\_prep.pro}.  Additional corrections were made for co-alignment between EIS's short wavelength (171--212\,{\AA}) and long wavelength (250--290\,{\AA}) bands  \cite{2008SoPh..248..457Y}; instrument and orbital jitter variations \cite{Shimizuetal2007}; the sinusoidal spectrum drift of the lines on the CCD due to orbital changes \cite{Mariskaetal2007PASJ}, and the tilt of the emission lines on the detector. These corrections resulted in an absolute wavelength calibration of $\pm$\,4.4 km s$^{-1}$ \cite{2010SoPh..266..209K}.

To investigate the magnetic structure of the events studied, we obtained closest-time data from the {\it Michelson Dopper Imager} (MDI; \opencite{Scherreretal1995SoPh}) instrument onboard the SOHO satellite. The full disk MDI line-of-sight (LOS) magnetograms possessed a pixel resolution of 1\farcs98 $\times$ 1\farcs98 and were obtained with a 96-min cadence over the roughly 12 h of EIS observations. MDI level-1.8 magnetogram data, which are the definitively calibrated images of the solar line-of-sight (LOS) photospheric magnetic field, were processed using the SSW routine {\sf  mdi\_prep.pro} for the removal of instrument noise. In order to increase the signal-to-noise ratio, three magnetograms were averaged at each time step.

\begin{table}[!ht]
\caption{List of EIS spectral emission lines used in the present study.}
\label{tbl:eis_elines}
\vspace{0.5mm}
\begin{tabular}{cccc}
\hline
Ion & Wavelength (\AA) & Transition & $\log T$ (K)\\
\hline
O \textsc{iv}    & 279.80 &
2s$^2$ 2p $^2$P$_{3/2}$ - 2s$^2$ 3s $^2$S$_{1/2}$ & 5.3 \\
O \textsc{v}     & 192.90 &
2s 2p $^3$P$_2$ - 2s 3d $^3$D$_3$ & 5.4 \\
Fe \textsc{viii} & 185.21 &
3p$^6$ 3d $^2$D$_{5/2}$ - 3p$^5$ 3d$^2$ $^2$F$_{7/2}$ & 5.6 \\
Mg \textsc{vii}  & 278.75 &
2s$^2$ 2p$^2$ $^3$P$_2$ - 2s 2p$^3$ $^3$S$^1$ & 5.8 \\
Fe \textsc{x}    & 184.30 &
3s$^2$ 3p$^5$ $^2$P$_{1/2}$ - 3s$^2$ 3p$^4$ ($^3$P) 3d $^2$P$_{3/2}$ & 6.0 \\
Fe \textsc{xi}   & 180.40 &
3s$^2$ 3p$^4$ $^3$P$_2$ - 3s$^2$ 3p3 ($^4$S) 3d $^3$D$_3$ & 6.1 \\
Fe \textsc{xii}  & 195.12 &
3s$^2$ 3p$^3$ $^4$S$_{3/2}$ - 3s$^2$ 3p$^2$ ($^3$P) 3d $^4$P$_{5/2}$ & 6.2 \\
Si \textsc{x}    & 257.14 &
2s$^2$ 2p$^3$ $^4$S$_{3/2}$ - 2s 2p$^4$ $^4$P$_{1/2}$ & 6.2 \\
Si \textsc{xi}   & 191.26 &
2s$^2$ 2p$^2$ $^3$P$_2$ - 2s 2p$^3$ $^3$S$_1$ & 6.3 \\
Fe \textsc{xv}   & 284.16 &
3s$^2$ $^1$S$_0$ - 3s 3p $^1$P$_1$ & 6.4 \\
\hline
\end{tabular}
\end{table}

\begin{figure}[ht!]
\centering
\includegraphics[scale=0.68]{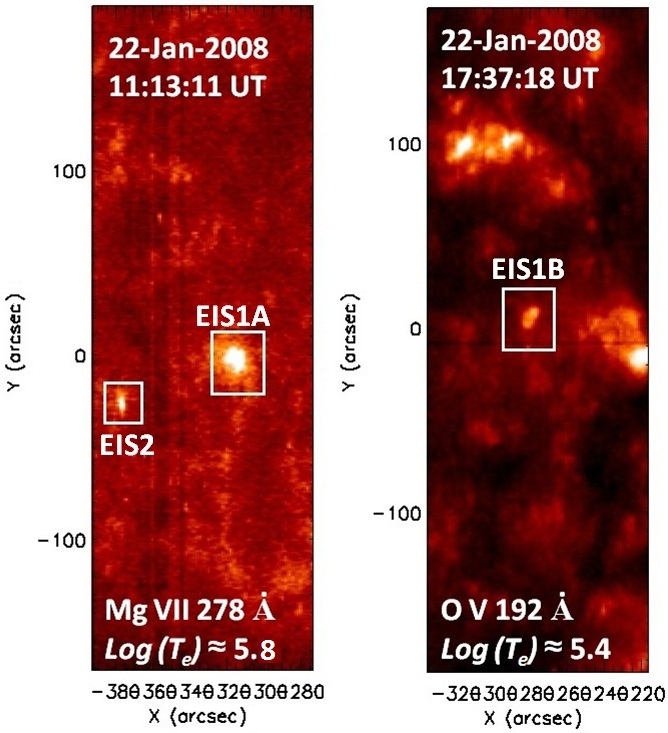}
   \caption{({\rm left}) EIS 2\arcsec FFOV intensity image (erg cm$^{-2}$ s$^{-1}$) in the emission of Mg\,\textsc{vii} $\lambda$278.39 ($\log T = 5.8$) observed on 22 January 2008 at 11:13:11 UT. The boxed regions indicate the locations of the EIS1A and EIS2 events. ({\rm right}) EIS 2\arcsec FOV intensity image (erg cm$^{-2}$ s$^{-1}$) of O\,\textsc{v} $\lambda$192.90 ($\log T = 5.4$) observed on 22 January 2008 at 17:37:18 UT, with boxed regions showing the EIS1B events.}
   \label{fig:EISraw}
\end{figure}

To co-align the EIS and MDI data, each of the ten EIS raster scans was mapped to solar coordinates by calibrating each scan's raster positions with respect to  Stonyhurst heliographic coordinates \cite{Zarro2005}. The magnetograms were then co-aligned to each EIS observation time by using the SSW routine {\sf drot\_map}, resulting in a final alignment error of $\lesssim$\,2\arcsec. This co-alignment error was measured by cross-correlating visually bright EUV coronal plasma structures to strong magnetogram regions. Projection effects were not expected to play a significant role in the co-alignment since each of our images were $\lesssim$\,$\pm$\,100\arcsec to disk center. Moreover, the observation time differences between the EIS scans and MDI magnetograms were $\lesssim$\,30 min.

Event selection was carried out via visual inspection of both coronal and TR images to identify compact and high contrast EUV brightenings. These events were then compared to MDI magnetograms. We selected three events. The first two events, which were an upper TR BP and a coronal BP (CBP), were required to exist in association with photospheric magnetic bipoles. The CBP had its peak temperature in the corona while the upper TR BP showed no discernable coronal emission. The third event, a blinker, was required to be associated with emerging photospheric magnetic flux. The upper TR BP and the coronal BP were both selected at 11:13:11 UT in the Mg\,\textsc{vii}~$\lambda$278.393\,\AA~($\log T = 5.8$) emission line. Figure~\ref{fig:EISraw} shows these events in two emission lines. The blinker was selected at 16:49:18 UT in the O\,\textsc{v}~$\lambda$192.900\,\AA~($\log T = 5.4$) emission line. After each event was selected their heliographic solar coordinates were then tracked throughout the EIS observation sequence by accounting for solar rotation ($\approx$\,0.17 arcsec min$^{-1}$) and instrument pointing \cite{SarroBerihuete2011}.

\begin{figure}[!t]
\begin{center}
  \includegraphics[scale=0.3]{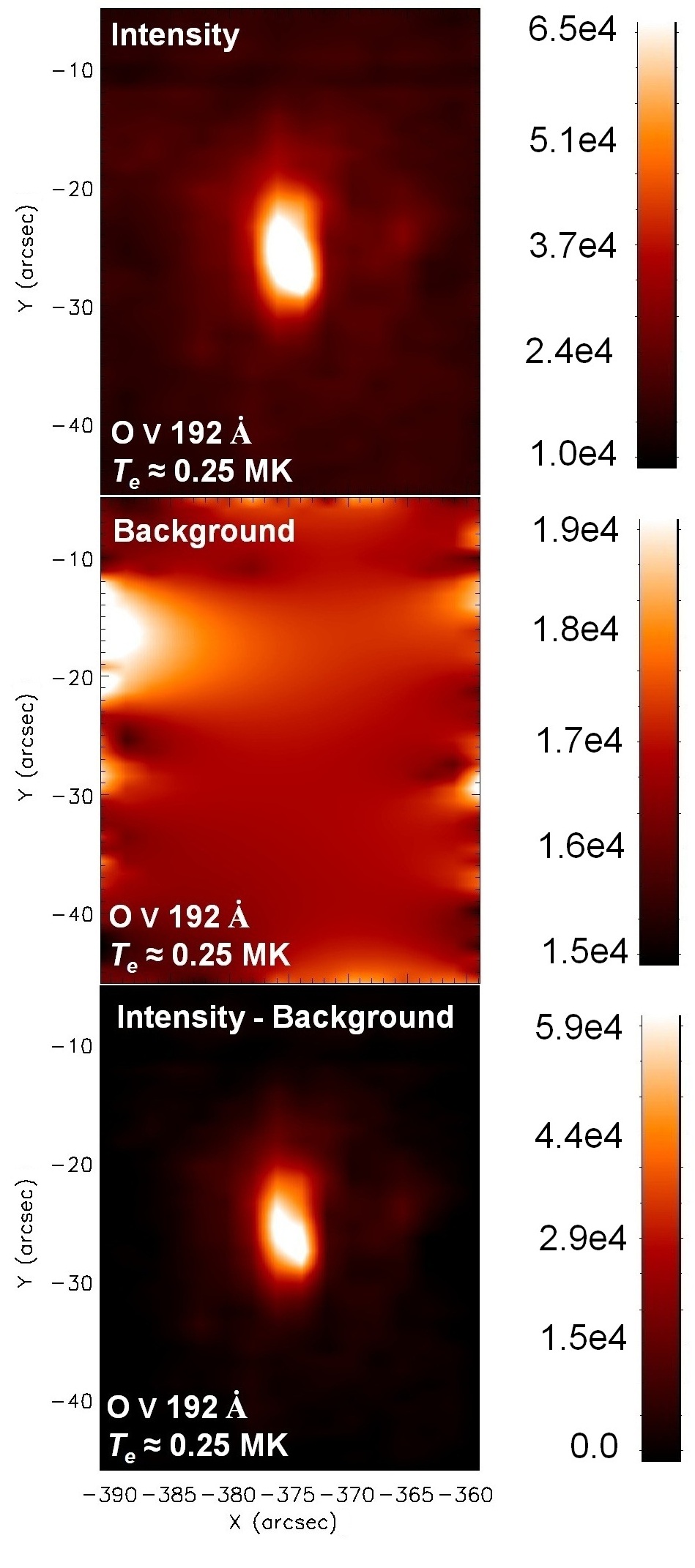}
   \caption{EIS 2\arcsec intensity (erg cm$^{-2}$ s$^{-1}$ sr$^{-1}$), background, and intensity minus background images (from top to bottom, respectively)  of O \textsc{v} (192.90\,{\AA}, $\log T = 5.4$) observed on 22 January 2008 at 11:13:11 UT of the EIS2 event studied here.}
\label{fig:bg_subtraction}
\end{center}
\end{figure}

\section{Measurements}
\label{sec:measurements}
For each of our three events we have fit the spectrum of each pixel with a Gaussian profile using the SSW routine {\sf eis\_auto\_fit.pro} to determine the line intensity at that pixel, the line center at that pixel from which Doppler velocities were derived, and the line width for each pixel from which non-thermal velocities were derived. We note that the O\,\textsc{v}\,$\lambda$192.90\,{\AA} line lies close to the EIS core line Ca\,\textsc{xvii}\,$\lambda$192.82\,{\AA}. \citeauthor{Youngetal2007PASJ} (\citeyear{Youngetal2007PASJ}) indicated that Ca\,\textsc{xvii} is blended with the Fe\,\textsc{xi}\,$\lambda$192.83\,{\AA} line which dominates in most solar conditions. Blending effects in O\,{\sc v} were determined to be minimal herein, which is consistent with suggestions that it comprises one half the emission seen at $\lambda$192.80\,{\AA} \cite{Youngetal2007PASJ}. Event morphology was measured from intensity images. The temperature structure of each event was examined via an emission measure (EM) loci analysis. Magnetic structure of events were determined through direct comparison of MDI line-of-sight magnetograms co-aligned with EUV EIS scans. Details of how each of these measurements were carried out is described in the subsections that follow.

\subsection{EUV Intensities and Lifetimes}
\label{sec:intensity}
Measurements of the integrated radiative flux and intensity enhancement of each event was obtained from the EIS intensity maps in each emission line. The first step in our process was to create a model of the background flux in each region of interest (ROI), which was a small rectangular region containing an event in its center with background/foreground emission in the surrounding pixels (Figure~\ref{fig:bg_subtraction}). This process was necessary since the observed intensity is the superposition of all emission along the line of sight through the optically thin plasma. The background was modeled by first performing a low-pass filter of the ROI in order to remove high-frequency noise such as bad pixels. Next, the border pixels of the ROI --- which contained only background/foreground emission ---  were held constant while the interior of the ROI was boxcar averaged using the standard IDL smoothing procedure. Care was taken to avoid overestimating the background or generating negative pixel values. The result of this procedure was the minimum area surface connecting the border pixels of the ROI. Figure~\ref{fig:bg_subtraction} shows an original image of the ROI, an image of the background model, and the background subtracted result. For our events, the background subtraction reduced pixel values by $\lesssim$ 15\%.

The integrated flux from each event at each time was calculated in two ways. First, the pixel values of the calibrated and background-subtracted ROIs were summed to obtain the flux $F_{\lambda}$ in each line. Second, the $3\sigma$ brightest pixels in each line were summed for each event to obtain $F^{3\sigma}_{\lambda}$. Finally, we calculated intensity enhancement $\psi_{\lambda}$ from the data prior to background subtraction using the equation:
\begin{equation}\label{eqn:ieh}
\psi_{\lambda} = \frac{F^{3\sigma; {\rm raw}}_{\lambda} - \bar{F}^{{\rm bg}}_{\lambda}}{\bar{F}^{{\rm bg}}_{\lambda}},
\end{equation}
where $F^{3\sigma; {\rm raw}}_{\lambda}$ is the average line flux from event pixels in each line $\lambda$ and $\bar{F}^{{\rm bg}}_{\lambda}$ is the mean solar background emission from each respective image. Uncertainties in $\psi_{\lambda}$ were then derived from photon counting statistics in both measurements of $F_{\lambda}$ and $\bar{F}^{{\rm bg}}_{\lambda}$.

Upon completing the measurements for all EIS observations we next identified which times the events existed in each line. Times with $\psi_{\lambda}$ $\ge$ 10\% were flagged as having an existing event. Lifetime measurements require existence continuity for two consecutive time-steps. Events that dropped below this threshold and reemerged later in the time sequence were considered different events given the $\approx$\,45 min interval between successive EIS observations.

\begin{figure}[!ht]
\begin{center}
  \includegraphics[scale=0.7]{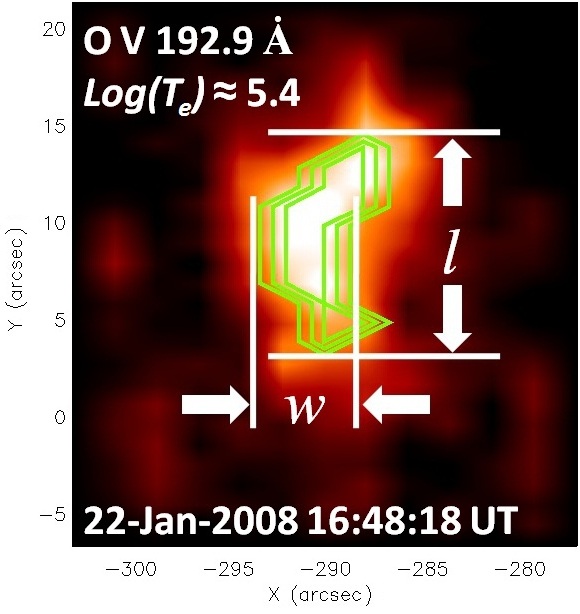}
   \caption{EIS 2\arcsec intensity image (erg cm$^{-2}$ s$^{-1}$ sr$^{-1}$) of O\,\textsc{v} $\lambda$192.90 ($\log T = 5.4$) observed 22 January 2008 at 16:49:18 UT of the EIS1b event. Green contours denote the event-unique pixels derived from our boundary extraction function.}
  \label{fig:er0b_O5_eups}
\end{center}
\end{figure}

\subsection{Morphologies and Doppler Velocities}
\label{sec:morphologies}
To measure the morphology of the event, we defined a set of event pixels per isolated thumbnail images. To define these pixel positions, we developed a boundary extraction function comprised of both downhill simplex and clumping techniques to isolate EUV emission enhancements from surrounding solar background. Our boundary extraction function assumes that a given image consists of a single EUV event in which significant contrast and the peak image flux exist. Events' morphologies are measured in each line using the largest spatial widths defined when our boundary extraction function is applied to the image. The greatest extent was labeled $l$ and the greatest width in the direction perpendicular to the greatest extent was labeled $w$ (Figure~\ref{fig:er0b_O5_eups}). Radiative areas $A_{\lambda}$ (km$^2$) were then derived from $l$ and $w$ assuming the telescope is at Lagrangian point 1 ($\approx$\,721 km arcsec$^{-1}$) and propagating uncertainties resultant from 1$\sigma$ Gaussian fits to the spatial dimensions corresponding to $l$ and $w$. These quantities were used with $F^{3\sigma}_{\lambda}$ to derive luminosities $P_{\lambda}$ (erg s$^{-1}$) of each event as a function of temperature. Luminosity uncertainties were propagated from photon counting statistics and axes length uncertainties.

The event pixels defined by our boundary extraction function were also used to measure, as a function of temperature, the characteristic LOS velocity $v_{\lambda}$ (km s$^{-1}$) for each event by averaging velocities returned by {\sf eis\_auto\_fit.pro} and propagating the 1$\sigma$ fit uncertainties.

\subsection{LOS Magnetic Field and Magnetic Flux}
\label{sec:magnetic}
For each event, the LOS magnetic field $|B_{\ell}|$ and flux $|\Phi|$ were measured. We also measured LOS field $|B_{\ell}|$ and flux $|\Phi|$ of positive and negative polarities. The sizes of the positive and negative polarities $A^{\pm}_{B}$ and separation distance $d_{B}$ were all measured. Note that the separation was not measured for the blinker event that was associated with a single unipolar region.

Signed fluxes were calculated by summing magnetogram pixels of each polarity. The unsigned flux was an integration of the absolute value of both positive or negative polarity fluxes. Magnetic polarity sizes, $A^{\pm}_{B}$, were obtained from the number of pixels used calculating signed fluxes. Measurements of separation distance $d_{B}$ were derived from calculations of the distance between the centroids of the positive and negative magnetic source elements. Uncertainties in LOS magnetic fields were derived from photon counting statistics on individual magnetogram pixels.

\subsection{Temperature Structure}
\label{sec:temperature}
To measure the amount of plasma emitting at a particular temperature we began with the standard definition of the column emission measure (EM),
%
\begin{equation}
{\rm EM} = \int_{S} n^{2}_{e} dh \hspace{.1in} \textrm{(cm$^{-5}$)},
\end{equation}
%
where $n_{{\rm e}}$ is the electron number density, $h$ is the LOS path length, and $S$ is the region containing the physically interesting plasma \cite{2011A&A...529A..21K}. Under the assumption that the observed plasma is optically thin and in ionization equilibrium, the observed intensity of a single emission line along the LOS $h$ is expressed as
%
\begin{equation}\label{eqn:synflux}
I_{\lambda} = A_{Z} \int_{S} G(T,n_{{\rm e}}) n^{2}_{{\rm e}} dh \hspace{.1in} \textrm{(erg cm$^{-2}$ s$^{-1}$ sr$^{-1}$)},
\end{equation}
%
where $A_{Z}$ is the elemental abundance, and $G(T,n_{{\rm e}})$ is the contribution function containing the relevant atomic parameters. Equation~(\ref{eqn:synflux}) can then be used to simplify the EM definition to the form
%
\begin{equation}
{\rm EM} = \frac{F_{\lambda}} {A_{Z} G(T,n_{{\rm e}})} \hspace{.1in} \textrm{(cm$^{-5}$)},
\end{equation}
%
which is the standard definition for deriving EM loci curves. These curves, calculated per observed emission line, represent the EM under the assumption that the measured EM originates from an isothermal plasma at a given temperature. As such, their usefulness lies in their ability to reveal isothermal plasmas where all curves meet at one temperature \cite{2011A&A...529A..21K}.

Prior to measuring EM loci curves, contribution functions were calculated from the CHIANTI atomic database (\opencite{Dereetal1997}, \citeyear{Dereetal2009}) by using the ionization equilibriums of \citeauthor{1992ApJ...398..394A} (\citeyear{1992ApJ...398..394A}), a typical quiet Sun electron density of 5 $\times$ 10$^{8}$ cm$^{-3}$ \cite{2010A&A...521A..21O}, and the photospheric abundances of \citeauthor{2007SSRv..130..105G} (\citeyear{2007SSRv..130..105G}). EM loci curves were then calculated for each event and all observation times by using measurements of $F_{\lambda}$ for each emission line in Table~\ref{tbl:EIS_data_tbl}.

\section{Results}
\label{sec:results}
Here we present the measurements for each event. As stated previously three events were identified: (a) an upper TR BP labeled EIS1a (lifetime, $\tau_{\lambda}$ $\approx$\,1.5 h), (b) a blinker labeled EIS1b ($\tau_{\lambda}$ $\approx$\,3.2 h), and (c) a CBP labeled EIS2 ($\tau_{\lambda}$ $\approx$\,0.8 h). Note that EIS1b occurred in the same ROI as EIS1a but appeared $\approx$\,4 h after EIS1a had disappeared. Table~\ref{tbl:EIS_data_tbl} presents the measurements of the three events. Below each event is discussed in turn.

\begin{table}[!h]
\caption{List of parameters derived for the three EIS events studied. ``Obs.time" means initial time of observation. Angular brackets indicate time averages.}
\label{tbl:EIS_data_tbl}
\vspace{0.5mm}
\begin{tabular}{lccccccc}
\hline
\hline
 & Obs.time & $\tau_{\lambda}$ & $\langle A \rangle$ & $\langle P \rangle$ & $\langle \psi \rangle$ & $\langle v \rangle$ \\
 &     & [h] & [km$^2$] & [erg s$^{-1}$] & [\%] & [km s$^{-1}$] \\
\hline
EIS1a & 11:13:11 UT & 1.5 & 9.3$\times$10$^{7}$ & 1.3$\times$10$^{22}$ & 128 & 2.90 \\
& & & 2.4$\times$10$^{8}$ & 1.2$\times$10$^{22}$ & 29 & 12.4 \\

EIS1b & 16:49:18 UT & 3.2 & 4.1$\times$10$^{8}$ & 9.3$\times$10$^{21}$ & 29 & 12.4 \\
& & & 1.1$\times$10$^{9}$ & 5.3$\times$10$^{22}$ & 11 & 47.1 \\
EIS2 & 11:13:11 UT & 0.8 & 1.2$\times$10$^{7}$ & 1.7$\times$10$^{21}$ & 209 & 17 \\
& & & 2.3$\times$10$^{7}$ & 5.6$\times$10$^{22}$ & 247 & 20.8 \\
\hline
\end{tabular}
\end{table}

\begin{figure}[!ht]
\begin{center}
  \includegraphics[scale=0.32]{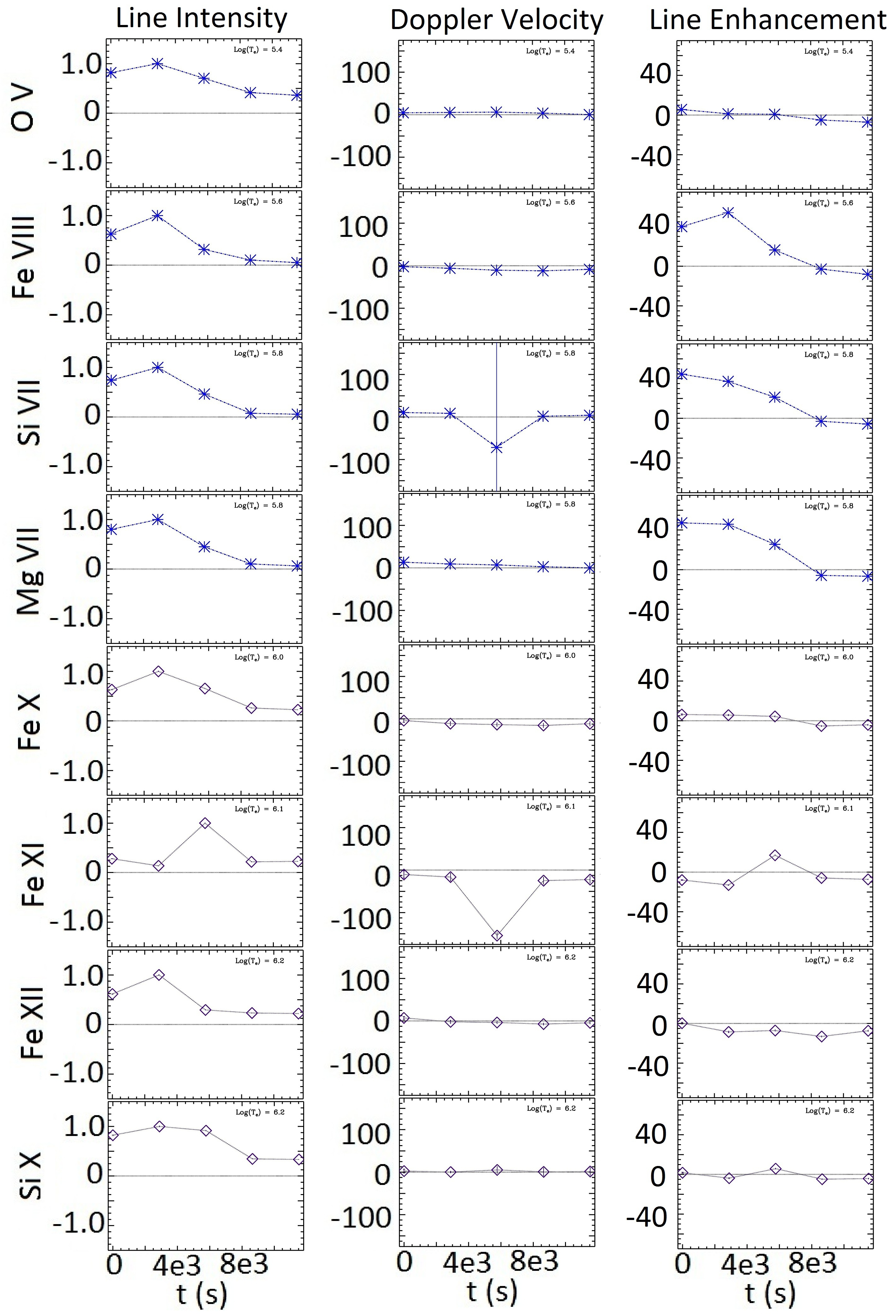}
   \caption{Integrated line fluxes $F_{\lambda}$ (arbitrary units), LOS velocities $v_{\lambda}$ (km s$^{-1}$), and emission enhancement $\psi_{\lambda}$ (\%) vs. time (s) in each of eight emission lines. The time is relative to 11:13:11 UT.}
   \label{fig:er0a_TR_lcs}
\end{center}
\end{figure}

\subsection{EIS1a --- TR Bright Point}

Figure~\ref{fig:er0a_TR_lcs} shows the temporal variation of the integrated line intensities $F_{\lambda}$ (arbitrary units), LOS velocities $v_{\lambda}$ (km s$^{-1}$), and emission enhancement $\psi_{\lambda}$ (\%) vs. time (s) in each of eight emission lines for the EIS1a event. The bulk of emission lines exhibited a significant increase in the integrated flux values $F_{\lambda}$ during the first 45 min and had either returned to or dropped below their initial values 45 min later. EUV intensity enhancements $\psi_{\lambda}$ were $\approx$\,40\% above the background for the TR lines but were negligible in the coronal lines. LOS velocities were minimal in comparison to the surrounding background ($\lesssim \pm 20$ km s$^{-1}$) and were close to the resolution limit of velocities ($\approx \pm 5$ km s$^{-1}$).

\begin{figure}[ht!]
\begin{center}
  \includegraphics[scale=0.2,angle=90]{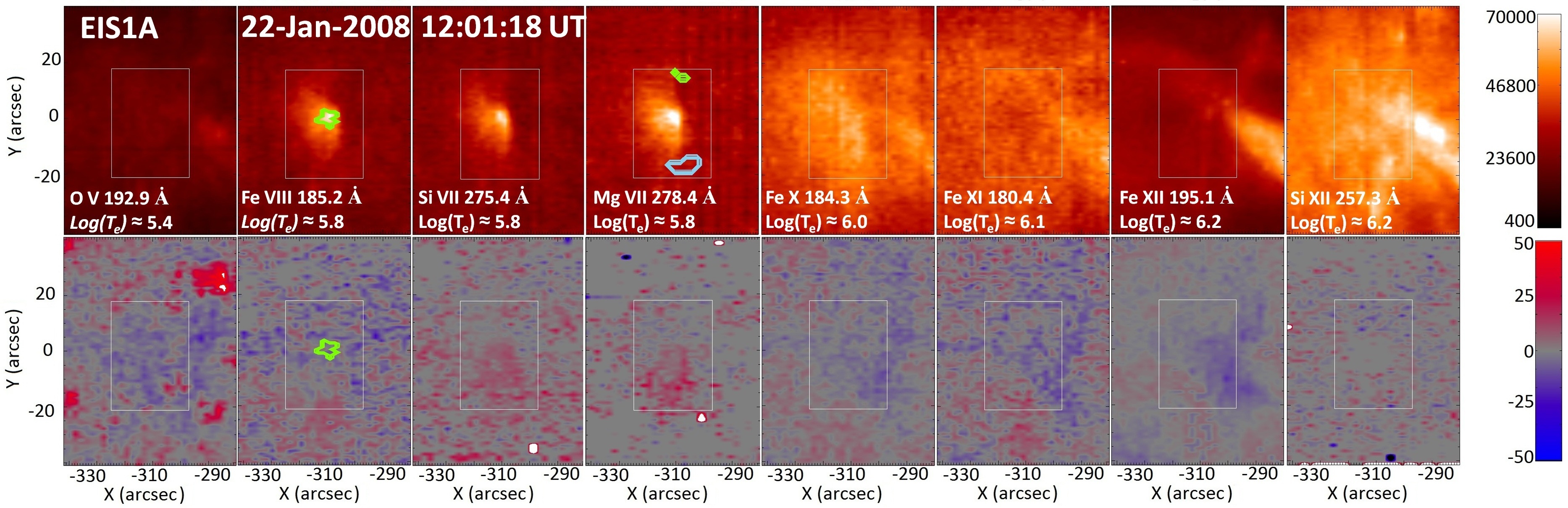}
   \caption{Intensity and LOS velocity maps for EIS1a at the peak emission time, 12:01:18 UT. The squares indicate the areas selected for its analyses. Green contours on the Fe\,\textsc{viii} intensity map denote the event pixels determined from our boundary extraction function. Green and blue contours on the Mg\,\textsc{vii} intensity image denote the locations of the positive and negative magnetic field patches.}
   \label{fig:er0a_2D_lcs}
\end{center}
\end{figure}

In Figure~\ref{fig:er0a_2D_lcs} we have provided EUV intensity maps, LOS velocity maps, and MDI magnetic field contours at the time of peak emission, 12:01:18 UT. These images illustrate that the TR BP was thermally isolated within the TR as observed by previous authors. This is most notabe at $\log T = 5.8$ as observed in the emission of Fe\,\textsc{viii} $\lambda$185.2, Si\,\textsc{vii} $\lambda$275.4, Mg\,\textsc{vii} $\lambda$278.4. At peak brightness, the size and luminosity of EIS1a in the TR was $\approx 9 \times 10^{7}$ km$^{2}$ and $\approx 1 \times 10^{22}$ erg s$^{-1}$, respectively. These values agree well with those reported for the TR BPs presented by \citeauthor{Bruni2006} (\citeyear{Bruni2006}) as observed by the SUMER instrument in the emission of Ne\,\textsc{viii} $\lambda$770.41 ($\log T = 5.8$). Likewise, the  TR intensity enhancements are outside of the range previously reported for blinkers. Moreover, whereas blinkers have been mostly identified as EUV enhancements in the O\,\textsc{v} emission line \cite{1999A&A...351.1115H}, the TR BP EIS1a showed no significant O\,\textsc{v} emission.

The locations of event pixels identified by our boundary extraction function, which were used in our measurement of the characteristic LOS velocity of the event, are shown as contours on the Fe\,\textsc{viii} $\lambda$185.2 intensity and velocity maps in Figure~\ref{fig:er0a_2D_lcs}. Although significant plasma flows were observed in the vicinity of EIS1a at the time of peak brightness, the pixels spatially coincident with EIS1a do not show a strong Doppler signal. The values observed are $\lesssim$ $\pm$ 20 km s$^{-1}$, which is consistent with Doppler flows reported for quiet region CBPs \cite{Madjarskaetal2003A&A,Tianetal2008ApJ,Perez-Suarez2008A&A}.

\begin{figure}[!ht]
\begin{center}
  \includegraphics[scale=0.62]{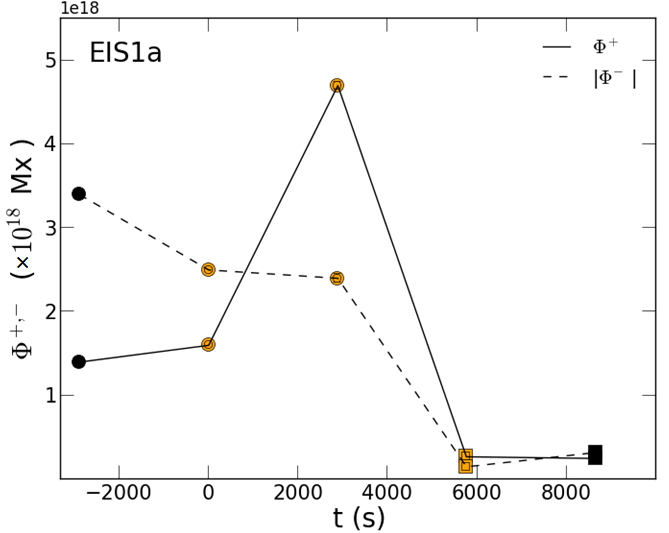}
   \caption{Positive (solid) and negative (dashed) magnetic fluxes ($\times 10^{18}$ Mx) as a function of time (s) for EIS1a. Square and circle symbols denote observations where the magnetic field configuration was fragmented and bipolar respectively, while orange marks observation times where the EUV event existed, $\approx$ 0 -- 6$\times$10$^3$ s. The time is relative to 11:13:11 UT. 
   }\label{fig:EIS_er0a_signedflux}
\end{center}
\end{figure}

Co-spatial magnetic field data shows an interacting bipolar pair separated by $\approx$\,25 Mm, as illustrated on the Mg\,\textsc{vii} intensity map as green and blue contours. Until its complete cancellation at 12:49:19 UT, the separation distance of the flux pair was increasing at a rate of $\approx 2 \times 10^{-2}$ Mm min$^{-1}$. In Figure~\ref{fig:EIS_er0a_signedflux} we present the temporal evolution of the positive and negative fluxes. At the event onset, 11:13:11 UT, the positive and negative fluxes were $\approx$\,$\pm$\,2$\times$10$^{18}$ Mx. Emerging positive flux occurred prior to the event's peak luminosity in EUV. During this peak the pair's typical positive and negative flux was $\approx 4 \times 10^{18}$ Mx. From the EUV flux peak till the event ceased to exist the bipolar pair canceled completely at a rate of $\approx 6 \times 10^{18}$ Mx h$^{-1}$ (Figure~\ref{fig:EIS_er0a_signedflux}).

The flux of the bipole is consistent with those expected in the TR \cite{2003ApJ...598.1387P} and values reported for TR BPs \cite{Bruni2006}, while observed cancellation rates are typical of CBP phenomena \cite{Leeetal2011ApJ,Zhangetal2012ApJ}. As expected, for BPs \cite{Leeetal2011ApJ,1997ApJ...491..952K,2001ApJ...553..429L}, both magnetic flux emergence and cancellation are found to be directly related to fluctuations of and brightening in solar EUV images. The flux pair's spatial divergence and magnetic flux evolution are considered as direct evidence of magnetic reconnection then.

Under the observed conditions either or both of the following can occur: formation of a null point above the flux pair as they come into contact which produces impulsive reconnection events \cite{1994ApJ...427..459P}, or a 2D-like current sheet forms along the fan-separatrix that is responsible for the long-lived nature of these brightenings \cite{Pariatetal2010ApJ}. Given the absence of any spectral signature of a vertically orientated jet \cite{Zhangetal2012ApJ}, we consider our results indicative of the latter scenario. The diverging flux elements observed in magnetograms $\approx 6$ h prior to our EIS observation sequence suggest that the flux pair forced the current sheet to an orientation that was highly angled from the normal to the solar surface, {\it i.e.}, more horizontally directed. This seems a likely explanation for the minimal coronal emission enhancements, given that such field geometries would lead to plasma motion with higher velocity components parallel to the solar surface \cite{Hegglandetal2009ApJ}. Such a scenario could explain the minimal enhancements of intensity images at temperatures cooler than $\log T = 5.4$, given the direction in which the heated plasma is propelled.

The TR intensity images, which shows emission ``smeared out" away from a bright central core, are also consistent with this scenario. It is in the direction of this smearing that TR LOS velocity showed enhanced downflow regions which in relation to our above proposed scenario would be indicative of cooling and falling plasma as it spread out from the bright core. Furthermore, the positive $x$ side of the event ({\it i.e.}, opposite side to the smearing) is visually dim and maybe be suggestive of plasma evaporation \cite{Hegglandetal2009ApJ} as a result of the reconnection events.

\begin{figure}[ht!]
\begin{center}
  \includegraphics[scale=0.63]{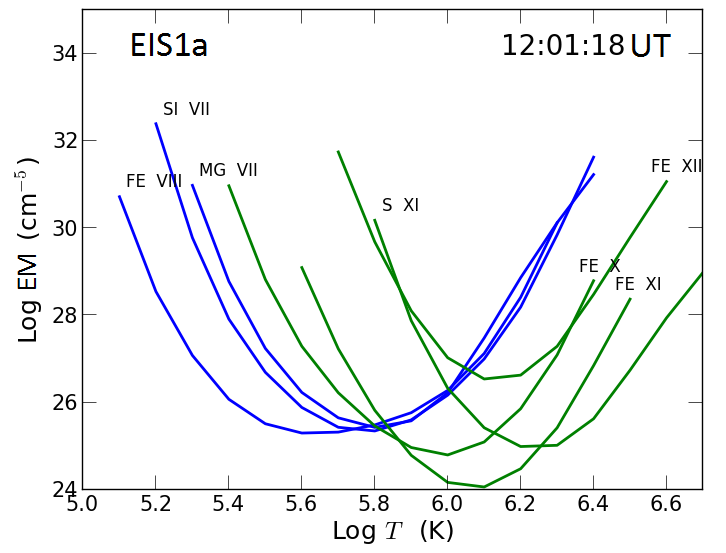}
   \caption{EM loci curves (blue and green represent TR and coronal emission lines, respectively) as a function of temperature for EIS1a at the time of peak enhancement (12:01:18 UT), showing a distinctive isothermal component around $\log\,(T)\,\approx\,5.8$.}
   \label{fig:er0a_pems_plocis}
\end{center}
\end{figure}

EIS1a's EM loci analysis indicates isothermal plasma is located at $\log T = 5.8$, where the convergence of multiple curves is observed (Figure~\ref{fig:er0a_pems_plocis}). This is consistent with its similar visual appearance  in emission lines sharing this formation temperature (the top row of Figure~\ref{fig:er0a_2D_lcs}). Higher TR EMs in comparison with coronal EMs  suggest condensation \cite{CraigMcClymont1986}.  Otherwise, the TR loci curve distribution is relatively broad (Figure~\ref{fig:er0b_emplocis}). This is perhaps indicative of unresolved structuring \cite{Brooksetal2012ApJ}. Considering the previous notion that TR BPs are a self-similar phenomena to CBPs, which have been observed to be composed of flaring loop systems \cite{1981SoPh6977H,Habbaletal1990ApJ,Koutchmyetal1997A&A,1997ApJ...488..499K,Tianetal2008ApJ}, one would expect the presence of both bi-directional flows
(\opencite{Athayetal1982ApJ}; \citeyear{Athayetal1983ApJb}; \opencite{Athayetal1983ApJa}; \opencite{Gebbieetal1981ApJ}) and unresolved structure \cite{Zhangetal2012ApJ}.

The cooling of coronal plasma as the source of bright EUV emission is eliminated by noting the observed upward flowing plasma, $\approx 15$ km s$^{-1}$, at $\log T = 6.0 - 6.2$. Moreover, these temperatures converge about the isothermal location. A more likely interpretation is that the conversion of magnetic to thermal energy took place over a small temperature regime, somewhere around $\log T = 5.8 - 5.9$. This is consistent with the bi-directional velocity pattern observed in LOS velocity images from the Fe\,\textsc{viii} emission line (Figure~\ref{fig:er0a_2D_lcs}), as well as suggestions by \citeauthor{Hegglandetal2009ApJ} (\citeyear{Hegglandetal2009ApJ}) that such characteristics can be used as probes for reconnection heights. When we compare our observation of very small LOS velocities to those typically expected in jets resulting from newly reconnected field lines \cite{Pariatetal2010ApJ,2011A&A...529A..21K}, this further supports our suggestion that the dominant component of the outflow velocities are likely parallel to the solar surface. Given the uncertainty in the formation temperature Fe\,\textsc{viii} \cite{DelZanna2012arXiv}, we only suggest the reconnection events are occurring at the upper TR temperature of Fe. The cooling of heated upper TR plasma explains both the presence of redshifts and similarly structured brightenings in emission lines with formation temperatures close to this range. Furthermore, these inferences are consistent with previous suggestions on the current sheet orientation as well as minimal coronal EUV emission enhancements.

\subsection{EIS1b --- Blinker}
\label{sec:blinker}
\begin{figure}[ht!]
\begin{center}
  \includegraphics[scale=0.32]{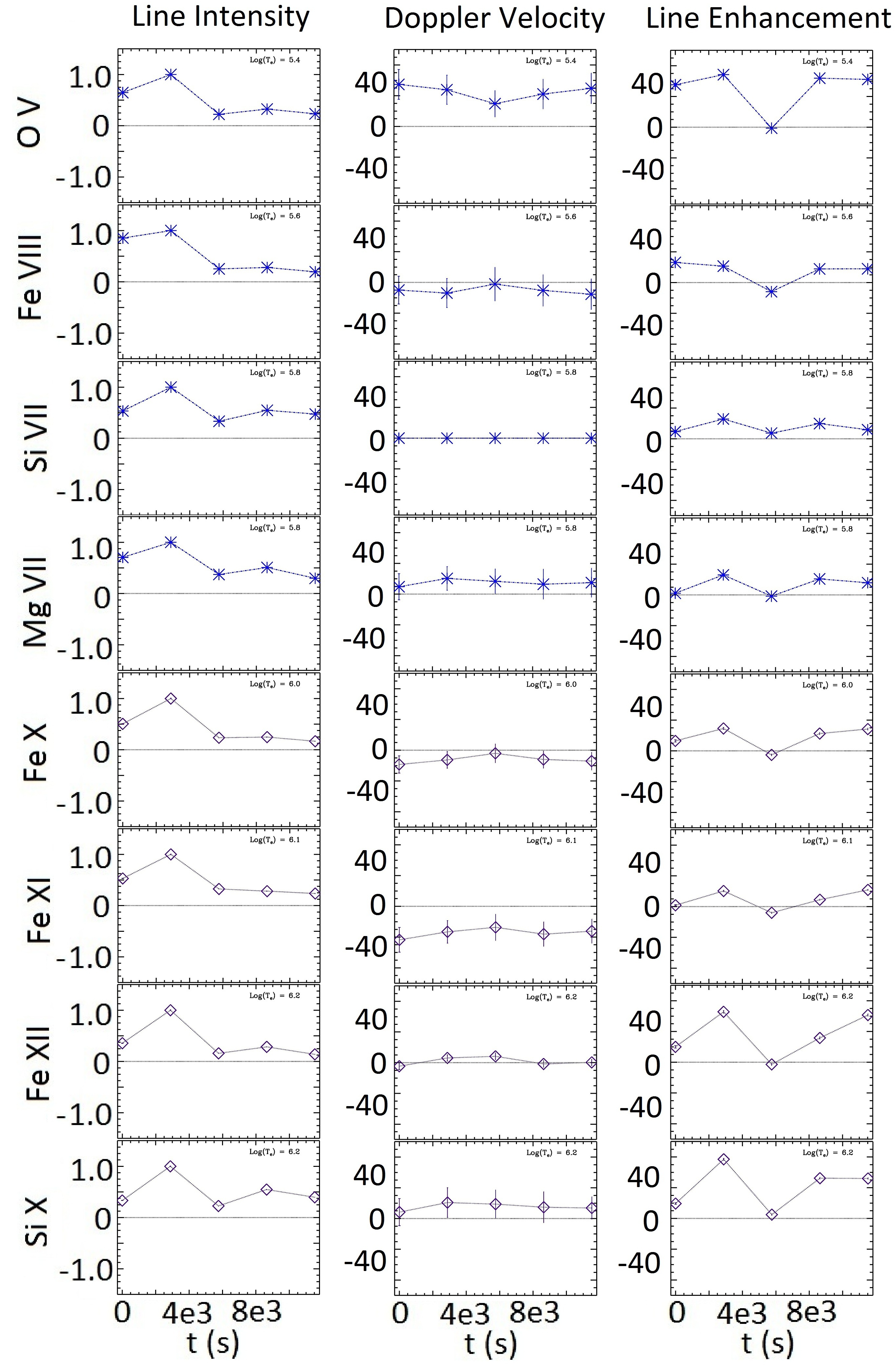}
   \caption{$F_\lambda$ (arbitrary units), LOS velocity (km s$^{-1}$), and emission enhancement (\%) vs. time (s) in columns from left to right, respectively, associated with EIS1b which existed over the 0 s -- 1$\times$10$^4$ s temporal domain. It is noted the time is relative to 16:01:19 UT.}
   \label{fig:er1b_TR_lcs}
\end{center}
\end{figure}

The temporal evolution of integrated flux $F_{\lambda}$, LOS velocity $v_{\lambda}$, and EUV emission enhancements $\psi_{\lambda}$ of the EIS1b event are provided in Figure~\ref{fig:er1b_TR_lcs}. Its flux evolution exhibits a distinctive complex light curve in $\approx 75\%$ of the emission lines, with a secondary event being triggered at the conclusion of the first. The intensities of this latter event are typically $\approx 50\%$ less than the first while disappearing at temperatures of $\log T = 6.0 - 6.3$. The events are similar in terms of their durations ($\approx 1.75$ h) and their heating and cooling timescales ($\approx 45$ min). The time delay between intensity peaks is $\approx 6 \times 10^{3}$ s. This period is consistent with with $\approx 8\%$ of blinkers observed by \citeauthor{1999A&A...351.1115H} (\citeyear{1999A&A...351.1115H}), while both non-complex and complex light curves have been attributed to this phenomena \cite{2004A&A...422..709B}.

The evolution of the EUV emission enhancement $\psi_{\lambda}$ was consistent with the changes in the integrated flux as a function of temperature for EIS1b, with the exception of the Mg~\textsc{vii} line whose temporal evolution was smooth and singly peaked (Figure~\ref{fig:er1b_TR_lcs}) and occurred simultaneously with the onset of the second event. No significant difference in the magnitudes of the successive $\psi_{\lambda}$ peaks was observed for this line. During this event, the O\,\textsc{v} $\lambda$192.9 ($\log T = 5.4$) peak intensity enhancement was $\approx 69\%$ while in the emission of Fe\,\textsc{x} $\lambda$184.30 ($\log T = 6.0$) it was $\approx 19\%$. These results, as well as the resultant TR to corona enhancement gradient, are typical of what has been observed for blinkers \cite{2003A&A...409..755H}.

LOS velocities were $\lesssim \pm 20$ km s$^{-1}$ except for the O\,\textsc{v} emission line which was $\approx 25$ km s$^{-1}$ on average. As observed in Figure~\ref{fig:er1b_TR_lcs}, the O\,\textsc{v} LOS velocity evolution correlates well with that of the radiative flux in that the increases in downward flowing plasma correlate with the occurrence of increases in the EUV emission. These TR flow speeds are consistent with those observed in correlation with both quiet and active region blinkers \cite{2003A&A...403..731M,Bewsheretal2003SoPh,Brooksetal2004ApJ}. At $\log T = 6.0$ and 6.1, this same correlation between LOS velocity and radiative flux is observed for upward flowing plasma. This is suggestive of an upper TR bi-directional jet \cite{Hegglandetal2009ApJ}.

\begin{figure}[!ht]
\begin{center}
  \includegraphics[scale=0.18,angle=90]{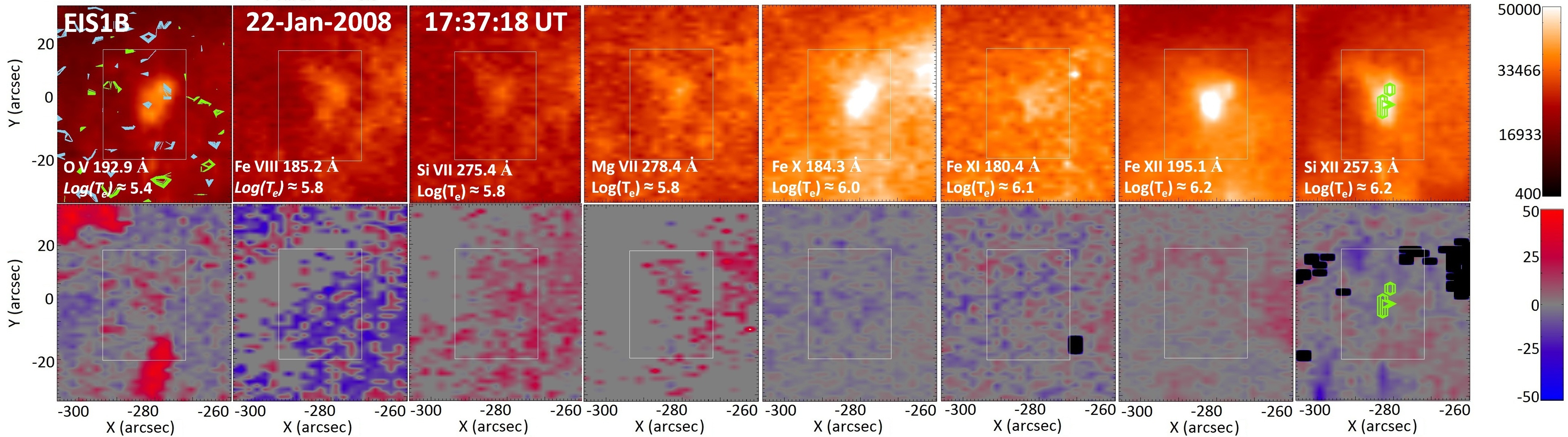}
   \caption{Intensity and LOS velocity maps for EIS1b at the peak emission time, 17:37:18 UT. The squares indicate the areas selected for its analysis. Green contours on the Si \textsc{x} intensity map denote the temperature regime in which they were defined and the spatial location identified as the event pixels. On the O \textsc{vii} intensity image green and blue contours represent positive and negative LOS magnetic field plotted at levels of $\pm$ 8 -- 15 G.}
   \label{fig:er0b_2D_lcs}
\end{center}
\end{figure}

Figure~\ref{fig:er0b_2D_lcs} provides intensity and LOS velocity maps as a function of temperature for EIS1b at 17:37:18 UT, which corresponds to the peak emission time of the first event. Visual inspection reveals a distinctly different morphology and intensity enhancement in the different emission lines. EIS1b was significantly dimmer for emission lines with peak formation temperatures of $\log T = 5.8$. In the TR the event was $\approx 4 \times 10^{8}$ km$^2$ in size with a luminosity of $\approx  9 \times 10^{21}$ erg s$^{-1}$. In the coronal lines it was an order of magnitude larger in both its projected area and its luminosity. Considering only a single heating and cooling event, the total energy radiated was $\approx 5 \times 10^{25}$ erg. Though the event's average size was three times larger than a typical blinker's, the resultant energy is consistent with the thermal energy estimates predicted by \citeauthor{1997AdSpR..20.2239H} (\citeyear{1997AdSpR..20.2239H}a,b) and observed by \citeauthor{1999A&A...351.1115H} (\citeyear{1999A&A...351.1115H}).

\begin{figure}[!t]
\begin{center}
  \includegraphics[scale=0.62]{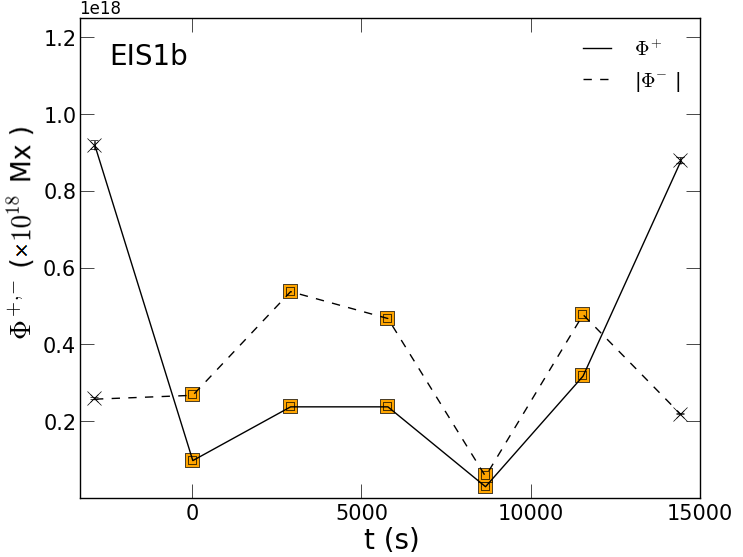}
   \caption{Positive (solid) and negative (dashed) magnetic fluxes ($\times 10^{18}$ Mx) as a function of time (s) for EIS1b. \textit{Square} and \textit{X} symbols denote observations where the magnetic field configuration was fragmented and unipolar respectively, while orange marks observation times where the EUV event existed, $\approx$ 0 -- 1.1$\times$10$^4$ s. The time is relative to 16:01:19 UT.}
   \label{fig:EIS_er1b_signedflux}
\end{center}
\end{figure}

Magnetograms co-spatial with EIS1b were dominated by fragmented flux. The evolution of the positive and negative flux is shown in Figure~\ref{fig:EIS_er1b_signedflux}. The absolute value of the negative flux is $<$\,3$\times$10$^{17}$ Mx throughout the observational sequence and was approximately three times greater than the positive flux. The dominant single polarity is typical of blinkers \cite{1999A&A...351.1115H,2002SoPh..206...21B} while the magnetic field strength is consistent with values reported for quiet region blinkers by \citeauthor{Bewsheretal2002ESASP} (2002). The first peak in EUV intensity correlates with emergence of both positive and negative flux of $\approx$\,1.4$\times$10$^{17}$ Mx and \,2.7$\times$10$^{17}$ Mx, respectively. For the secondary event, the peak emission correlates with magnetic cancellation occurring at rates of $\approx$\,4.3$\times$10$^{15}$ Mx min$^{-1}$ and 8.6$\times$10$^{15}$ Mx min$^{-1}$ for positive and negative flux, respectively.

In relationship to their EUV brightness, both the emergence and cancellation events are equivalent. However, the emergence event correlates with significantly higher EUV flux rates at both TR and coronal heights. These results indicate there exists no correlation between peaks in EUV brightness and the strength/evolution of the underlying magnetic field. Moreover, they are consistent with the conclusions that blinkers are a direct result of density enhancements by reconnection events by previous authors (\citeauthor{Bewsheretal2002ESASP}, 2002; \opencite{2003A&A...409..755H}).

\begin{figure}[ht!]
\begin{center}
  \includegraphics[scale=0.6]{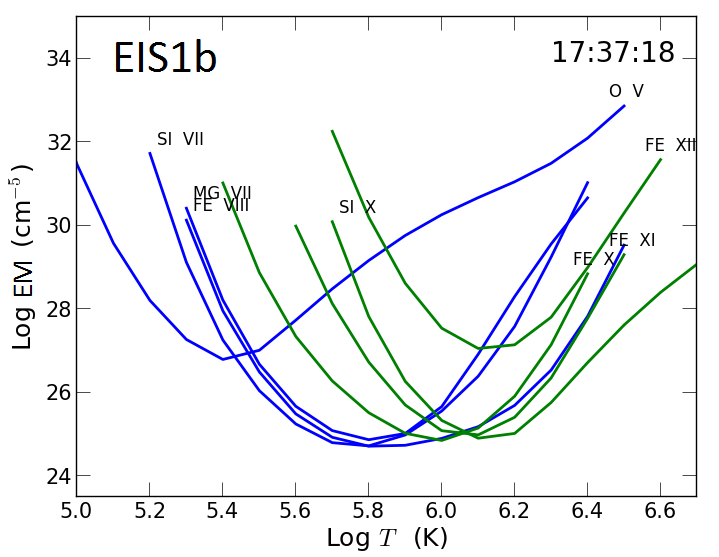}
   \caption{EM loci curves (blue and green represent TR and coronal emission lines, respectively) as a function of temperature for EIS1b at 12:01:18 UT, showing a mildly isothermal component around $\log\,(T)\,\approx\,6.1$.}
   \label{fig:er0b_emplocis}
\end{center}
\end{figure}

Our EM loci analysis of EIS1b indicates weak isothermal components at $\log T = 6.1$ for observation times corresponding to peaks in emission. The EM loci curves of the first peak are shown in Figure~\ref{fig:er0b_emplocis}. The absence of convergence for the cooler TR loci curves, as compared to those of the corona, indicates only the hotter coronal plasma is isothermal. Broad and approximately constant TR EM distributions, most notably at $\log T = 5.8$, are an indication of unresolved structure \cite{Brooksetal2012ApJ}, which is consistent with the visual structuring of the event in intensity images (Figure~\ref{fig:er0b_2D_lcs}).

The lower TR emission measure, compared to that of the corona, has been related to chromospheric evaporation \cite{CraigMcClymont1986,McClymontCraig1987}. Under this interpretation coronal blue shifts indicate the heights to which chromospheric plasma is being heated, while pervasive TR redshifts are explained as a result of the cooling and condensing of the heated coronal plasma. This interpretation is preferred over heating from a bi-directional jet, as stated earlier, given that the magnetic field analysis indicated only a casual relationship between reconnection and EUV flux increases. We recognize that blinkers do not typically consist of isothermal plasma, at least in the TR, based on the fact that they are related to density-enhancements \cite{2003A&A...409..755H}. However, under our proposed scenario it seems likely that the TR brightenings are due to density enhancements being generated via plasma heated to coronal temperatures.

\subsection{EIS2 --- Coronal Bright Point}
\label{sec:cbp}
\begin{figure}[ht!]
\begin{center}
  \includegraphics[scale=0.32]{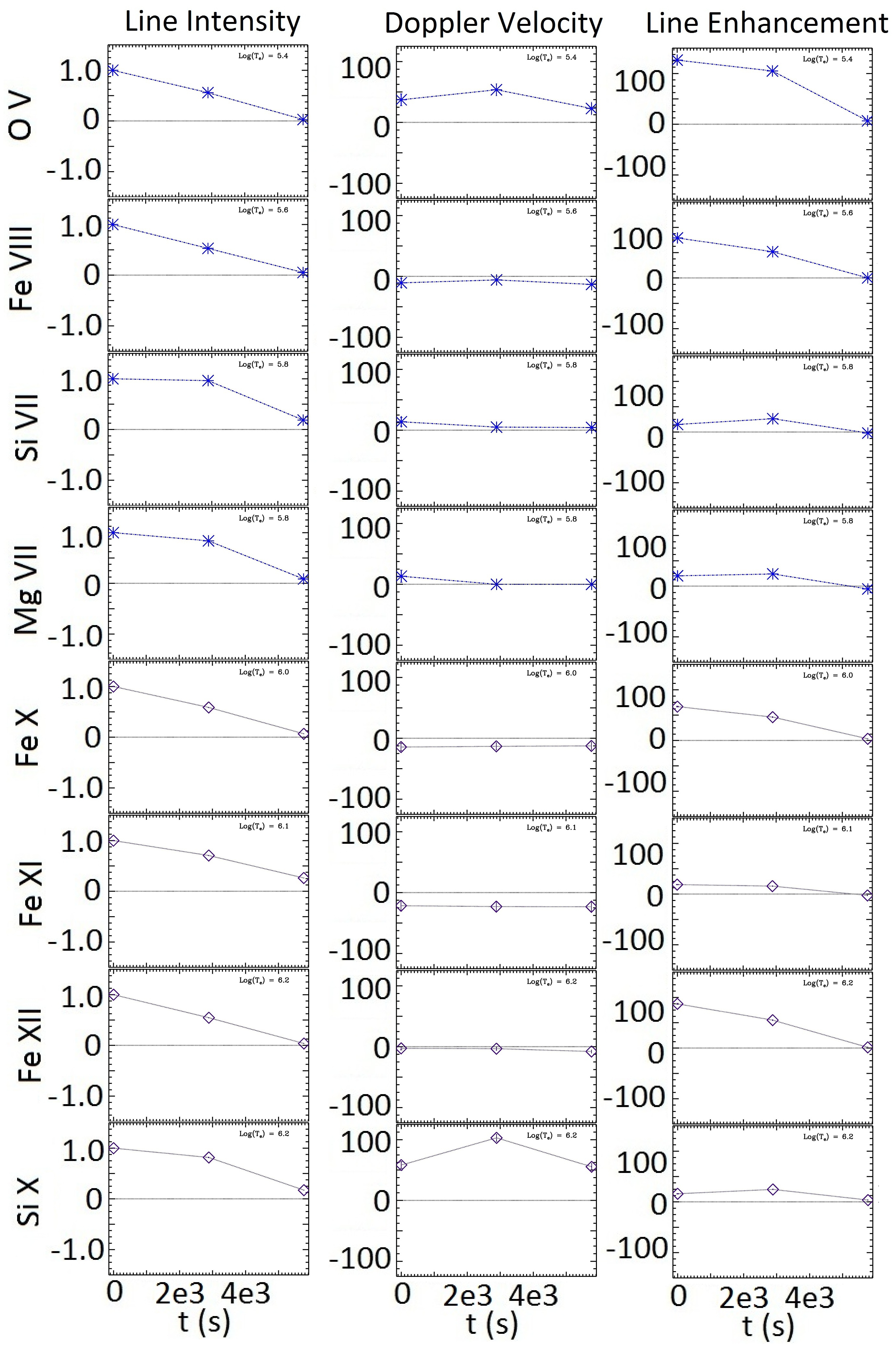}
   \caption{$F_\lambda$ (arbitrary units), LOS velocities (km s$^{-1}$), and emission enhancement (\%) vs. time (s) for EIS2 in columns from left to right, respectively. The times quoted are relative to 11:13:11 UT.}\label{fig:eis2_TR_lcs}
\end{center}
\end{figure}

The temporal evolution of the integrated flux $F_{\lambda}$, LOS velocities $v_{\lambda}$, and EUV emission enhancements $\psi_{\lambda}$ of the EIS2 event are provided in Figure~\ref{fig:eis2_TR_lcs}. Decreases in radiative flux occurred independent of temperature over the entire range of temperatures probed by our observations, {\it i.e.}, from $\log T = 5.4$ to 6.2. This implies that our temporal sequence witnessed only the cooling phase of this event or that the cadence of our EIS observations was longer than the typical heating timescale. Based on the evolution of the emission enhancements, which showed increases or remained constant during the event's evolution, we have settled on the latter interpretation. Further support of this interpretation is found by considering that the brightness increases occurred on a timescale of $\approx$ 45 min, which are consistent with the observed CBP brightness oscillation periods \cite{Tianetal2008ApJ}. Also our observed coronal emission enhancements ($\approx$ 2.5 times that of the background) agree well with those reported for quiet region CBPs \cite{Zhangetal2012ApJ}.

As with other events studied here, EIS2 is associated with very small LOS velocities at the bulk of temperature regimes studied. The exceptions are in the emission of O\,\textsc{v} $\lambda$192.90 at $\log T = 5.4$ and in the emission of Si\,\textsc{x} $\lambda$274.14 at $\log T = 6.2$, where the plasma was falling at a rate of $\approx$ 50 km s$^{-1}$. TR plasma downflow speeds are within the ranges observed in microflares \cite{2009A&A...505..811B}, while coronal flows are consistent with \citeauthor{2011A&A...529A..21K}'s (\citeyear{2011A&A...529A..21K}) results for microflarring CBPs. It is noted that these flow speed peaks correlate in time with peaks in the intensity enhancements $\psi_{\lambda}$ of the respective emission lines (Figure~\ref{fig:eis2_TR_lcs}).

\begin{figure}[!ht]
\begin{center}
  \includegraphics[scale=0.20,angle=90]{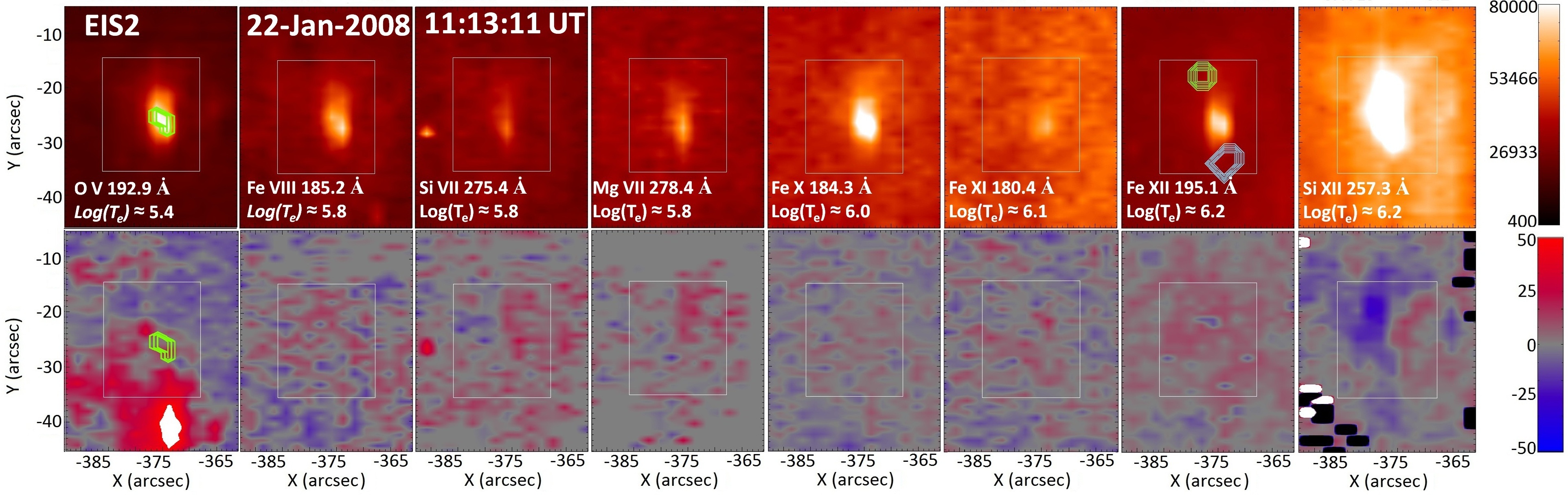}
   \caption{Intensity and LOS velocity maps for EIS2 at the peak emission time, 11:13:11 UT. The squares indicate the areas selected for its analysis. Green contours on the O \textsc{v} intensity map denote the temperature regime in which they were defined and the spatial location identified as the event pixels. On the Fe \textsc{xii} intensity image green and blue contours denote the locations of the positive and negative magnetic patches.}
   \label{fig:er1a_2D_lcs}
\end{center}
\end{figure}

We have provided both intensity and LOS velocity images in Figure~\ref{fig:er1a_2D_lcs} for the time of EIS2's peak irradiance. The intensity maps are consistent with previous reports of multi-thermal emission across the TR and corona for CBPs \cite{1997ApJ...491..952K,Kwonetal2012ApJ}. EIS2's typical size was $\approx 1 - 2\times 10^{7}$ km$^{2}$ in both the TR and corona. Its luminosity ranged from $\approx 10^{21}$ erg s$^{-1}$ to $10^{22}$ erg s$^{-1}$ for the TR and corona, respectively. These sizes and lumnosities are very similar to those previously reported for CBPs by other authors \cite{2003ApJ...598.1387P,2001ApJ...553..429L}.

The LOS velocity results shown in Figure~\ref{fig:eis2_TR_lcs} are consistent with the visual nature observed in Figure~\ref{fig:er1a_2D_lcs}. Inspection of LOS velocity maps in Figure~\ref{fig:er1a_2D_lcs} reveal a distinctive bi-directional jet for the Si \textsc{x} ($\log T \approx 6.2$) emission line. EIS2 occurred co-spatially with the downflow jet ($\lesssim 50$ km s$^{-1}$) whose upward motion reached speeds of $\gtrsim 50$ km s$^{-1}$. This is expected since dimming and brightenings are most typically associated with blue-shifted and red-shifted regions, respectively \cite{CraigMcClymont1986,McClymontCraig1987}. These flow speeds are consistent with \citeauthor{2011A&A...529A..21K}'s (\citeyear{2011A&A...529A..21K}) study, but are direct evidence that diffuse coronal jets can occur in conjunction with quiet region CBPs.

\begin{figure}[!ht]
\begin{center}
  \includegraphics[scale=0.62]{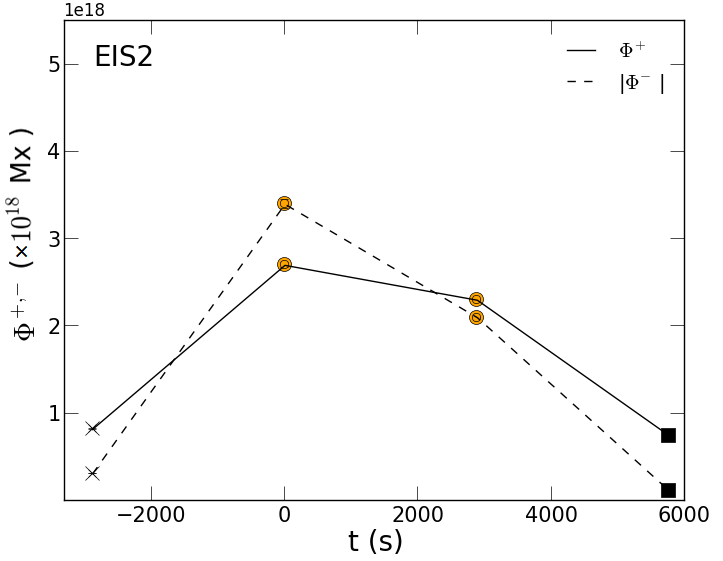}
   \caption{Positive (solid) and negative (dashed) magnetic fluxes ($\times 10^{18}$ Mx) as a function of time (s) for EIS2. Square, circle, and cross symbols denote observations where the magnetic field configuration was fragmented bipolar, and unipolar respectively, while orange marks observation times where the EUV event existed, $\approx$ 0 -- 2.5$\times$10$^3$ s. }
   \label{fig:EIS_er1a_signedflux}
\end{center}
\end{figure}

At 11:13:11 UT EIS2 occurred co-spatial with a bipolar pair in magnetograms, separated by $\approx$ 12 Mm. The evolution of positive and negative fluxes is provided in Figure~\ref{fig:EIS_er1a_signedflux}. During the 45 min time span in which the EUV event existed, the bipolar pair's flux remained relatively constant (Figure~\ref{fig:EIS_er1a_signedflux}), while the separation distance decreased by $\approx 40\%$. The typical magnetic flux associated with the bipolar pair was $\approx 3 \times 10^{18}$ Mx. Ultimately, the EUV event disappeared in conjunction with the complete cancellation of the photospheric bipole. We found the  typical flux cancellation rate, of both positive and negative flux, to be $\approx 1 \times 10^{18}$ Mx h$^{-1}$.

EIS2's magnetic flux cancellation rate is consistent with those reported by \citeauthor{Leeetal2011ApJ} (\citeyear{Leeetal2011ApJ}), while its longer time scale indicates that the resultant brightness increases are not explained by slow mode waves trapped within the magnetic flux tubes constituting the structure \cite{Doyleetal1998SoPh,Ugarte-Urraetal2004A&A}. Based on similarities in magnetic field changes between our event and those reported by \citeauthor{Leeetal2011ApJ} (\citeyear{Leeetal2011ApJ}) and \citeauthor{Zhangetal2012ApJ} (\citeyear{Zhangetal2012ApJ}), we suggest that magnetic reconnection initiated by the convergence of the bipolar pair was responsible for the brightening observed in the EUV data. Moreover, these reconnection events were recurrent given that the typical timescale for coronal reconnection is tens of minutes \cite{Chenetal1999ApJ}. In this proposed scenario the solar surface motions form a current sheet and build up magnetic energy near a null point of the assumed small-scale loop system comprising the CBP \cite{Zhangetal2012ApJ}. Thus, successive and impulsive reconnection events may be triggered at the current sheet site until the flux pair has canceled completely \cite{Pariatetal2010ApJ}.

\begin{figure}[ht!]
\begin{center}
  \includegraphics[scale=0.65]{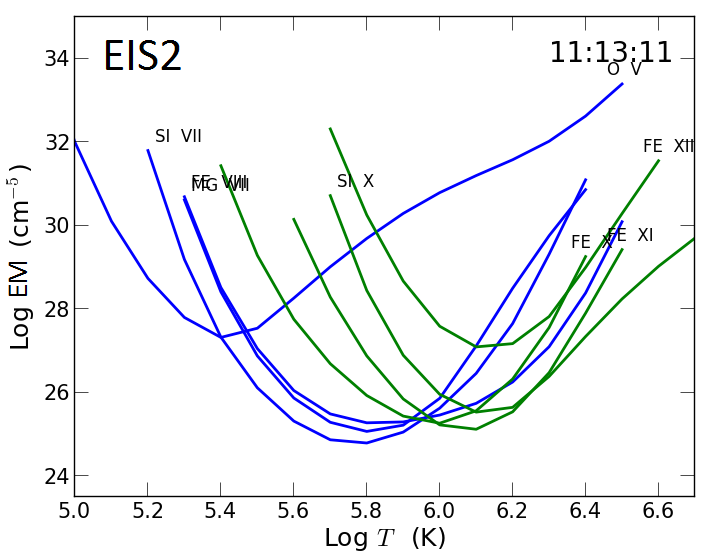}
   \caption{EM loci curves (blue and green represent TR and coronal emission lines, respectively) as a function of temperature for EIS2 at 12:01:18 UT showing a distinctive isothermal component around $\log\,(T)\,\approx\,6.0$, and weaker secondary component around $\log\,(T)\,\approx\,6.2$.}
   \label{fig:er1a_emplocis}
\end{center}
\end{figure}

Isothermal analysis of EIS2 as shown in Figure~\ref{fig:er1a_emplocis} indicated that the event consisted of isothermal plasma around $\log T = 5.9$ at 11:13:11 UT. The occurrence of EM loci curves reaching higher than this location are indicative of an additional isothermal component \cite{2011A&A...529A..21K}. In Figure~\ref{fig:er1a_emplocis} both the O\,\textsc{v} and Fe\,\textsc{xii} EM loci curves are distributed significantly higher than the remaining emission lines, while intensity images of these lines share a distinctly similar bright EUV structure (Figure~\ref{fig:er1a_2D_lcs}). From these results we hypothesize that O\,\textsc{v} was ionized at temperatures more typical of the corona. Therefore, the additional isothermal component is likely at $\log T \approx 6.2$.

\section{Discussion}
\label{sec:discussion}
We have presented observations of a TR BP, blinker, and CBP, each recorded on 22 January 2008 during the 11:13:11 UT to 23:12:40 UT time frame, from multiple EIS emission lines along with accompanying MDI LOS magnetic field data. Our work has presented the first measurements of Doppler signatures associated with TR BPs as well as the first analyses of their detailed physical characteristics in multiple emission lines. With these data we may now identify the similarities and differences between the TR BPs and the CBP and blinker.

Our three events share a common characteristic in terms of their EUV morphology; they are all spatially compact and unresolved brightenings with characteristic sizes of $\approx 10^{7}$ km$^{2}$ -- $10^{8}$ km$^{2}$ (Figures~\ref{fig:er0a_2D_lcs}, \ref{fig:er0b_2D_lcs}, and \ref{fig:er1a_2D_lcs}). As well, the luminosities of each of the three events were consistent with the luminosity versus unsigned magnetic flux relation of \citeauthor{2003ApJ...598.1387P} (\citeyear{2003ApJ...598.1387P}). We do note, however, that the emission enhancements of the BPs were more than four times larger than for the blinker.

The photospheric magnetic signatures of the TR BP and the CBP events were distinctly different from that of the blinker. The TR BP occurred on the neutral line of an interacting magnetic bipolar pair, suggesting that magnetic reconnection involving the said pair was responsible for its generation. This suggestion is supported by the appearance of the TR BP in EUV when the bipolar pair emerged followed by the observation that the EUV from the TR BP became undetectable after the bipolar pair was completely canceled. This behavior suggests that the generation mechanism of the TR BP involved reconnection between the newly emerged bipole and surrounding background field, given the observed decreases in both positive and negative magnetic flux over its lifetime (Figure~\ref{fig:EIS_er0a_signedflux}). Moreover, it is this characteristic that leads us to rule out the idea that this event is a direct result of slow magnetic reconnection which occurs as new flux emerges. The resultant heating energy of $\approx 10^{25}$ erg is consistent with nanoflares that impulsively heat the plasma during nanoflare``storms". This result supports the idea that like CBPs, TR BPs can be considered flaring loop systems where individual strands or small bundles are being rapidly heated and cooled at differing rates. Unfortunately, we were unable to resolve the heating time scales of the individual events and thus cannot directly infer the rate at which magnetic energy is converted to thermal energy via individual reconnection events. Nonetheless, the self-similar generation of the TR BP and CBP are an indication that TR and coronal heating are not fundamentally different for bright point phenomena. Moreover, it provides further support that the cooler BPs are a manifestation of multi-scaled self-similarity occurring in the solar atmosphere.

The connection between quiet region microflares and CBPs has been established \cite{2011A&A...529A..21K}, but only those occurring in coronal hole regions have been linked with jets. Our spectral analysis of CBPs, mainly the bi-directional jet observed at $\log T$\,$\approx$\,6.2, indicates that jets may also exist in quiet regions, too. Thereby, we have provided the first observational evidence of the association between CBPs and jets in the quiet Sun. However, the difference in the large scale magnetic configuration yields a fundamental difference in the coronal hole and quiet region jets. That is, the coronal hole jets eject material nearly normal to the solar surface. This is consistent with the coronal hole events occurring in regions where the large-scale field is open \cite{2011A&A...529A..21K}. Our quiet Sun events appear to eject material in a direction that is nearly parallel to the solar surface. In essence, the plasma heating, radiation generation, and plasma ejection processes may be similar for TR BPs and CBPs. However, it is the large scale field configuration that determines the fate of any ejected material and whether the material is accelerated away from the Sun thus contributing to the solar wind flux, or whether the material falls back to the solar surface. The latter events are similar to explosive events (\opencite{Beckers1968SoPh}).

The single polarity dominated magnetic fields above which the blinker occurred are consistent with the idea that these events are generated by slow magnetic reconnection with pre-existing fields. We also observed significant emergence and cancellation events of the dominate polarity field occurring prior to the appearance of the said event, which is expected under the previous description of blinkers. We speculate, based on spectral results which showed no significant mass upflows, that this reconnection likely took place in the corona given the strong downflows found therein. However, the increases in this speed within the TR ($T$ $\approx$ 0.25 MK) are puzzling and are better described by a scenario in which multiple atmospheric locations are experiencing the said reconnection. We do recognize that the secondary event is fit by the description where reconnection only took place at coronal heights.

In terms of the applicability of isothermal assumption, only the blinker was non-isothermal, consistent with ideas that density/filling factor enhancements are responsible for their generation \cite{2003A&A...409..755H}. This description is also supported by the spectral data, which indicate significant downward flowing plasma that temporally correlated with peaks in EUV flux. These flows are interpreted as mass transfer from the corona to cooler portions of the atmosphere. The decreasing plasma speeds with decreasing temperature lead us to hypothesize that coronal energy was transported downward possibly via magnetoacoustic waves. Once these waves encounter the steep density gradient from the corona to the TR they steepen into shocks thereby heating the surrounding plasma. It is recognized that for this suggestion to be plausible for both the first and the second blinker events, such shocks would be required to initiate reconnection in the TR and propel mass downwards.

The loci plots of the TR BP and CBP show that the latter event was more resolved based on the visual inspection of the resultant EM distribution widths. We hypothesize that the TR BP was heated impulsively from below based on its EM loci and the form of its light curve. Minimal evidence was found for the existence of unresolved structure for this event, which indicates that our temporal cadence was unable to resolve the heating time scale. The TR BP shared no energetic connection with the corona which gives support to our earlier suggestion that its origin is most likely attributed to in-situ heating. We discount this emission event as a direct result of cooling coronal material by noting the absence of significant mass-flows throughout the coronal temperatures. However, it is recognized that our slower temporal cadence could have resulted in missed peaks in coronal emission line intensity. For the CBP a significantly greater portion of the plasma at the isothermal component was radiating. However, at this time the two identified isothermal components are more likely attributed to this event being a microflarring CBP, which could also just as likely explain its more spatially resolved nature.

The deluge of solar observational data from near-constant monitoring of the Sun is driving the need for frameworks that can quickly and automatically produce classifications of observed sources. Recent development of modern machine learning techniques and high performance computing architectures have made possible the efficient execution of automated probabilistic multi-class classification of very large datasets in reasonable time frames
(\opencite{2007A&A...475.1159D}; \opencite{2009A&A...494..739S}, \citeyear{2009A&A...506..535S}; \opencite{2009A&A...506..519D}; \opencite{2011ApJ...733...10R}, \citeyear{2012ApJ...744..192R};  \opencite{2011MNRAS.418...96B};  \opencite{2012AJ....143..123M}; \opencite{2012ApJ...746..170M}; \opencite{2012PASP..124..280L}). An essential step in the development of such a framework is processing a large set of events of known class to train the classifier. The work presented here has indicated that similar measurements made across multiple event classes, {\it i.e.}, TR BP, blinker, and CBP, hold class-predictive power. Thereby, our work can be used in conjunction with similar studies \cite{2000ApJ...528L.119C,2003A&A...409..755H,2003A&A...403..731M,2004A&A...422..709B,2005A&A...432..307B,2009ApJ...701..253M} to begin facilitating a pathway which ultimately works towards automated multi-class classification of compact transients occurring in all temperature regimes of the solar atmosphere. As noted previously, this will be a major challenge of solar physics in the coming years. Finally it is emphasized that, the resultant large number statistics on a diversity of example structures across many classes from such a framework will allow a reliable understanding of the underlying physical processes of these events to be achieved.

\section{Conclusions}

An investigation of a TR BP, blinker, and CBP, observed using {\it Hinode}/EIS, was carried out in this paper. For the first time, we have presented results on Doppler signatures of TR BPs. Similarities shared and distinct observational differences between these three events were highlighted. The relationship each event shared with the evolution of the underlying magnetic field was also studied and discussed. We concluded that the TR BP and CBP were a manifestation of solar atmospheric multi-scaled self-similarity, and that the heating of TR and coronal plasma were not fundamentally different in the presence of these phenomena. A connection between quiet region CBPs and jets was established, with indications that the said jets occur at high look angles compared to the solar surface due to the large scale magnetic configuration of the quiet Sun.

Finally, our work has shown that similar measurements across multiple event classes hold class-distinguishing capabilities, and as such are a significant step towards automated multi-class classification of compact transients occurring throughout the solar atmosphere. Further investigations are required on such events as those analyzed herein to resolve their heating timescales and determine conclusively if the origin of their bright TR emission is coronal or direct evidence of in situ heating.

\section{Acknowledgements}
\label{sec:acknowledgements}
The authors greatly appreciate the referee's constructive comments and suggestions on the paper. {\it Hinode} is a Japanese mission developed and launched by ISAS/JAXA, with NAOJ as domestic partner and NASA and STFC (UK) as international partners. It is operated by these agencies in cooperation with ESA and NSC (Norway). This research was supported by National Aeronautics and Space Administration (NASA) grant NNX-07AT01G. This material is also based upon work supported by the National Science Foundation under Grant No. AST-0736479. The final publication is available at http://link.springer.com/article/
10.1007\%2Fs11207-013-0423-4.

\bibliographystyle{spr-mp-sola}
\bibliography{ubp1}

\begin{thebibliography}{94}
\ifx \bisbn   \undefined \def \bisbn  #1{ISBN #1}\fi
\ifx \binits  \undefined \def \binits#1{#1}\fi
\ifx \bauthor  \undefined \def \bauthor#1{#1}\fi
\ifx \batitle  \undefined \def \batitle#1{#1}\fi
\ifx \bjtitle  \undefined \def \bjtitle#1{\textit{#1}}\fi
\ifx \bvolume  \undefined \def \bvolume#1{\textbf{#1}}\fi
\ifx \byear  \undefined \def \byear#1{#1}\fi
\ifx \bissue  \undefined \def \bissue#1{#1}\fi
\ifx \bfpage  \undefined \def \bfpage#1{#1}\fi
\ifx \blpage  \undefined \def \blpage #1{#1}\fi
\ifx \burl  \undefined \def \burl#1{\textsf{#1}}\fi
\ifx \href  \undefined \def \href#1#2{\textsf{#2}}\fi
\ifx \doiurl  \undefined \def
  \doiurl#1{\href{http://dx.doi.org/#1}{\textsf{#1}}}\fi
\ifx \betal  \undefined \def \betal{\textit{et al.}}\fi
\ifx \binstitute  \undefined \def \binstitute#1{#1}\fi
\ifx \bctitle  \undefined \def \bctitle#1{#1}\fi
\ifx \beditor  \undefined \def \beditor#1{#1}\fi
\ifx \bpublisher  \undefined \def \bpublisher#1{#1}\fi
\ifx \bbtitle  \undefined \def \bbtitle#1{\textit{#1}}\fi
\ifx \bedition  \undefined \def \bedition#1{#1}\fi
\ifx \bseriesno  \undefined \def \bseriesno#1{\textbf{#1}}\fi
\ifx \blocation  \undefined \def \blocation#1{#1}\fi
\ifx \bsertitle  \undefined \def \bsertitle#1{\textit{#1}}\fi
\ifx \bsnm \undefined \def \bsnm#1{#1}\fi
\ifx \bsuffix \undefined \def \bsuffix#1{#1}\fi
\ifx \bparticle \undefined \def \bparticle#1{#1}\fi
\ifx \barticle \undefined \def \barticle#1{}\fi
\ifx \botherref \undefined \def \botherref#1{}\fi
\ifx \url \undefined \def \url#1{\textsf{#1}}\fi
\ifx \bchapter \undefined \def \bchapter#1{}\fi
\ifx \bbook \undefined \def \bbook#1{}\fi
\ifx \bcomment \undefined \def \bcomment#1{#1}\fi
\ifx \oauthor \undefined \def \oauthor#1{#1}\fi
\ifx \citeauthoryear \undefined \def \citeauthoryear#1{#1}\fi
\def \endbibitem {}
\ifx \bconflocation  \undefined \def \bconflocation#1{#1} \fi

\bibitem[\protect\citeauthoryear{{Abramenko}
  \textit{et~al.}}{2010}]{Abramenkoetal2010ApJ}
\begin{barticle}
\bauthor{\bsnm{{Abramenko}}, \binits{V.}},
\bauthor{\bsnm{{Yurchyshyn}}, \binits{V.}},
\bauthor{\bsnm{{Goode}}, \binits{P.}},
\bauthor{\bsnm{{Kilcik}}, \binits{A.}}:
\byear{2010},
\batitle{{Statistical distribution of size and lifetime of bright points
  observed with the New Solar Telescope}}.
\bjtitle{\apjl}
\bvolume{725},
\bfpage{L101}\,--\,\blpage{L105}.
doi:\doiurl{10.1088/2041-8205/725/1/L101}.
\end{barticle}
\endbibitem

\bibitem[\protect\citeauthoryear{{Arnaud} and
  {Raymond}}{1992}]{1992ApJ...398..394A}
\begin{barticle}
\bauthor{\bsnm{{Arnaud}}, \binits{M.}},
\bauthor{\bsnm{{Raymond}}, \binits{J.}}:
\byear{1992},
\batitle{{Iron ionization and recombination rates and ionization equilibrium}}.
\bjtitle{\apj}
\bvolume{398},
\bfpage{394}\,--\,\blpage{406}.
doi:\doiurl{10.1086/171864}.
\end{barticle}
\endbibitem

\bibitem[\protect\citeauthoryear{{Athay}, {Gurman}, and
  {Henze}}{1983}]{Athayetal1983ApJa}
\begin{barticle}
\bauthor{\bsnm{{Athay}}, \binits{R.G.}},
\bauthor{\bsnm{{Gurman}}, \binits{J.B.}},
\bauthor{\bsnm{{Henze}}, \binits{W.}}:
\byear{1983},
\batitle{{Fluid motions in the solar chromosphere-corona transition region. III
  - Active region flows from wide slit Dopplergrams}}.
\bjtitle{\apj}
\bvolume{269},
\bfpage{706}\,--\,\blpage{714}.
doi:\doiurl{10.1086/161080}.
\end{barticle}
\endbibitem

\bibitem[\protect\citeauthoryear{{Athay}
  \textit{et~al.}}{1982}]{Athayetal1982ApJ}
\begin{barticle}
\bauthor{\bsnm{{Athay}}, \binits{R.G.}},
\bauthor{\bsnm{{Gurman}}, \binits{J.B.}},
\bauthor{\bsnm{{Shine}}, \binits{R.A.}},
\bauthor{\bsnm{{Henze}}, \binits{W.}}:
\byear{1982},
\batitle{{Fluid motions the solar chromosphere-corona transition region. II
  Active region flows in C IV from narrow slit Dopplergrams}}.
\bjtitle{\apj}
\bvolume{261},
\bfpage{684}\,--\,\blpage{699}.
doi:\doiurl{10.1086/160379}.
\end{barticle}
\endbibitem

\bibitem[\protect\citeauthoryear{{Athay}
  \textit{et~al.}}{1983}]{Athayetal1983ApJb}
\begin{barticle}
\bauthor{\bsnm{{Athay}}, \binits{R.G.}},
\bauthor{\bsnm{{Gurman}}, \binits{J.B.}},
\bauthor{\bsnm{{Henze}}, \binits{W.}},
\bauthor{\bsnm{{Shine}}, \binits{R.A.}}:
\byear{1983},
\batitle{{Fluid motions in the solar chromosphere-corona transition region. I -
  Line widths and Doppler shifts for C IV}}.
\bjtitle{\apj}
\bvolume{265},
\bfpage{519}\,--\,\blpage{529}.
doi:\doiurl{10.1086/160695}.
\end{barticle}
\endbibitem

\bibitem[\protect\citeauthoryear{{Beckers}}{1968}]{Beckers1968SoPh}
\begin{barticle}
\bauthor{\bsnm{{Beckers}}, \binits{J.M.}}:
\byear{1968},
\batitle{{Solar spicules (invited review paper)}}.
\bjtitle{\solphys}
\bvolume{3},
\bfpage{367}\,--\,\blpage{433}.
doi:\doiurl{10.1007/BF00171614}.
\end{barticle}
\endbibitem

\bibitem[\protect\citeauthoryear{{Berkebile-Stoiser}
  \textit{et~al.}}{2009}]{2009A&A...505..811B}
\begin{barticle}
\bauthor{\bsnm{{Berkebile-Stoiser}}, \binits{S.}},
\bauthor{\bsnm{{G{\"o}m{\"o}ry}}, \binits{P.}},
\bauthor{\bsnm{{Veronig}}, \binits{A.M.}},
\bauthor{\bsnm{{Ryb{\'a}k}}, \binits{J.}},
\bauthor{\bsnm{{S{\"u}tterlin}}, \binits{P.}}:
\byear{2009},
\batitle{{Multi-wavelength fine structure and mass flows in solar
  microflares}}.
\bjtitle{\aap}
\bvolume{505},
\bfpage{811}\,--\,\blpage{823}.
doi:\doiurl{10.1051/0004-6361/200912100}.
\end{barticle}
\endbibitem

\bibitem[\protect\citeauthoryear{{Bewsher}, {Parnell}, and
  {Harrison}}{2002}]{2002SoPh..206...21B}
\begin{barticle}
\bauthor{\bsnm{{Bewsher}}, \binits{D.}},
\bauthor{\bsnm{{Parnell}}, \binits{C.E.}},
\bauthor{\bsnm{{Harrison}}, \binits{R.A.}}:
\byear{2002},
\batitle{{Transition region blinkers I. Quiet-Sun properties}}.
\bjtitle{\solphys}
\bvolume{206},
\bfpage{21}\,--\,\blpage{43}.
doi:\doiurl{10.1023/A:1014954629349}.
\end{barticle}
\endbibitem

\bibitem[\protect\citeauthoryear{{Bewsher}
  \textit{et~al.}}{}]{Bewsheretal2002ESASP}
\begin{botherref}
\oauthor{\bsnm{{Bewsher}}, \binits{D.}},
\oauthor{\bsnm{{Parnell}}, \binits{C.E.}},
\oauthor{\bsnm{{Brown}}, \binits{D.S.}},
\oauthor{\bsnm{{Hood}}, \binits{A.W.}}:
{Magnetic structure of transition region blinkers}.
In: {Sawaya-Lacoste}, H. (ed.)
\textit{SOLMAG 2002. Proceedings of the Magnetic Coupling of the Solar
  Atmosphere Euroconference, ESASP-505},
239\,--\,242.
\end{botherref}
\endbibitem

\bibitem[\protect\citeauthoryear{{Bewsher}
  \textit{et~al.}}{2003}]{Bewsheretal2003SoPh}
\begin{barticle}
\bauthor{\bsnm{{Bewsher}}, \binits{D.}},
\bauthor{\bsnm{{Parnell}}, \binits{C.E.}},
\bauthor{\bsnm{{Pike}}, \binits{C.D.}},
\bauthor{\bsnm{{Harrison}}, \binits{R.A.}}:
\byear{2003},
\batitle{{Dynamics of blinkers}}.
\bjtitle{\solphys}
\bvolume{215},
\bfpage{217}\,--\,\blpage{237}.
doi:\doiurl{10.1023/A:1025642612049}.
\end{barticle}
\endbibitem

\bibitem[\protect\citeauthoryear{{Bewsher}
  \textit{et~al.}}{2005}]{2005A&A...432..307B}
\begin{barticle}
\bauthor{\bsnm{{Bewsher}}, \binits{D.}},
\bauthor{\bsnm{{Innes}}, \binits{D.E.}},
\bauthor{\bsnm{{Parnell}}, \binits{C.E.}},
\bauthor{\bsnm{{Brown}}, \binits{D.S.}}:
\byear{2005},
\batitle{{Comparison of blinkers and explosive events: A case study}}.
\bjtitle{\aap}
\bvolume{432},
\bfpage{307}\,--\,\blpage{317}.
doi:\doiurl{10.1051/0004-6361:20041171}.
\end{barticle}
\endbibitem

\bibitem[\protect\citeauthoryear{{Blomme}
  \textit{et~al.}}{2011}]{2011MNRAS.418...96B}
\begin{barticle}
\bauthor{\bsnm{{Blomme}}, \binits{J.}},
\bauthor{\bsnm{{Sarro}}, \binits{L.M.}},
\bauthor{\bsnm{{O'Donovan}}, \binits{F.T.}},
\bauthor{\bsnm{{Debosscher}}, \binits{J.}},
\bauthor{\bsnm{{Brown}}, \binits{T.}},
\bauthor{\bsnm{{Lopez}}, \binits{M.}},
\bauthor{\bsnm{{Dubath}}, \binits{P.}},
\bauthor{\bsnm{{Rimoldini}}, \binits{L.}},
\bauthor{\bsnm{{Charbonneau}}, \binits{D.}},
\bauthor{\bsnm{{Dunham}}, \binits{E.}},
\bauthor{\bsnm{{Mandushev}}, \binits{G.}},
\bauthor{\bsnm{{Ciardi}}, \binits{D.R.}},
\bauthor{\bsnm{{De Ridder}}, \binits{J.}},
\bauthor{\bsnm{{Aerts}}, \binits{C.}}:
\byear{2011},
\batitle{{Improved methodology for the automated classification of periodic
  variable stars}}.
\bjtitle{\mnras}
\bvolume{418},
\bfpage{96}\,--\,\blpage{106}.
doi:\doiurl{10.1111/j.1365-2966.2011.19466.x}.
\end{barticle}
\endbibitem

\bibitem[\protect\citeauthoryear{{Brkovi{\'c}} and
  {Peter}}{2004}]{2004A&A...422..709B}
\begin{barticle}
\bauthor{\bsnm{{Brkovi{\'c}}}, \binits{A.}},
\bauthor{\bsnm{{Peter}}, \binits{H.}}:
\byear{2004},
\batitle{{Statistical comparison of transition region blinkers and explosive
  events}}.
\bjtitle{\aap}
\bvolume{422},
\bfpage{709}\,--\,\blpage{716}.
doi:\doiurl{10.1051/0004-6361:20040479}.
\end{barticle}
\endbibitem

\bibitem[\protect\citeauthoryear{{Brkovi{\'c}}, {Solanki}, and
  {R{\"u}edi}}{2001}]{2001A&A...373.1056B}
\begin{barticle}
\bauthor{\bsnm{{Brkovi{\'c}}}, \binits{A.}},
\bauthor{\bsnm{{Solanki}}, \binits{S.K.}},
\bauthor{\bsnm{{R{\"u}edi}}, \binits{I.}}:
\byear{2001},
\batitle{{Analysis of blinkers and EUV brightenings in the quiet Sun observed
  with CDS}}.
\bjtitle{\aap}
\bvolume{373},
\bfpage{1056}\,--\,\blpage{1072}.
doi:\doiurl{10.1051/0004-6361:20010652}.
\end{barticle}
\endbibitem

\bibitem[\protect\citeauthoryear{{Brooks}, {Warren}, and
  {Ugarte-Urra}}{2012}]{Brooksetal2012ApJ}
\begin{barticle}
\bauthor{\bsnm{{Brooks}}, \binits{D.H.}},
\bauthor{\bsnm{{Warren}}, \binits{H.P.}},
\bauthor{\bsnm{{Ugarte-Urra}}, \binits{I.}}:
\byear{2012},
\batitle{{Solar coronal loops resolved by Hinode and the Solar Dynamics
  Observatory}}.
\bjtitle{\apjl}
\bvolume{755},
\bfpage{L33}.
doi:\doiurl{10.1088/2041-8205/755/2/L33}.
\end{barticle}
\endbibitem

\bibitem[\protect\citeauthoryear{{Brooks}
  \textit{et~al.}}{2004}]{Brooksetal2004ApJ}
\begin{barticle}
\bauthor{\bsnm{{Brooks}}, \binits{D.H.}},
\bauthor{\bsnm{{Kurokawa}}, \binits{H.}},
\bauthor{\bsnm{{Kamio}}, \binits{S.}},
\bauthor{\bsnm{{Fludra}}, \binits{A.}},
\bauthor{\bsnm{{Ishii}}, \binits{T.T.}},
\bauthor{\bsnm{{Kitai}}, \binits{R.}},
\bauthor{\bsnm{{Kozu}}, \binits{H.}},
\bauthor{\bsnm{{Ueno}}, \binits{S.}},
\bauthor{\bsnm{{Yoshimura}}, \binits{K.}}:
\byear{2004},
\batitle{{Short-duration active region brightenings observed in the extreme
  ultraviolet and H{$\alpha$} by the Solar and Heliospheric Observatory Coronal
  Diagnostic Spectrometer and Hida Domeless Solar Telescope}}.
\bjtitle{\apj}
\bvolume{602},
\bfpage{1051}\,--\,\blpage{1062}.
doi:\doiurl{10.1086/381089}.
\end{barticle}
\endbibitem

\bibitem[\protect\citeauthoryear{{Bruni}}{2006}]{Bruni2006}
\begin{botherref}
\oauthor{\bsnm{{Bruni}}, \binits{D.}}:
2006,
{A comprehensive study of bright points in the sun's upper transition region}.
Master's thesis,
The University of Alabama in Huntsville.
\end{botherref}
\endbibitem

\bibitem[\protect\citeauthoryear{{Chae}
  \textit{et~al.}}{2000}]{2000ApJ...528L.119C}
\begin{barticle}
\bauthor{\bsnm{{Chae}}, \binits{J.}},
\bauthor{\bsnm{{Wang}}, \binits{H.}},
\bauthor{\bsnm{{Goode}}, \binits{P.R.}},
\bauthor{\bsnm{{Fludra}}, \binits{A.}},
\bauthor{\bsnm{{Sch{\"u}hle}}, \binits{U.}}:
\byear{2000},
\batitle{{Comparison of transient network brightenings and explosive events in
  the solar transition region}}.
\bjtitle{\apjl}
\bvolume{528},
\bfpage{L119}\,--\,\blpage{L122}.
doi:\doiurl{10.1086/312434}.
\end{barticle}
\endbibitem

\bibitem[\protect\citeauthoryear{{Chen}
  \textit{et~al.}}{1999}]{Chenetal1999ApJ}
\begin{barticle}
\bauthor{\bsnm{{Chen}}, \binits{P.F.}},
\bauthor{\bsnm{{Fang}}, \binits{C.}},
\bauthor{\bsnm{{Tang}}, \binits{Y.H.}},
\bauthor{\bsnm{{Ding}}, \binits{M.D.}}:
\byear{1999},
\batitle{{Simulation of magnetic reconnection with heat conduction}}.
\bjtitle{\apj}
\bvolume{513},
\bfpage{516}\,--\,\blpage{523}.
doi:\doiurl{10.1086/306823}.
\end{barticle}
\endbibitem

\bibitem[\protect\citeauthoryear{{Craig} and
  {McClymont}}{1986}]{CraigMcClymont1986}
\begin{barticle}
\bauthor{\bsnm{{Craig}}, \binits{I.J.D.}},
\bauthor{\bsnm{{McClymont}}, \binits{A.N.}}:
\byear{1986},
\batitle{{Quasi-steady mass flows in coronal loops}}.
\bjtitle{\apj}
\bvolume{307},
\bfpage{367}\,--\,\blpage{380}.
doi:\doiurl{10.1086/164423}.
\end{barticle}
\endbibitem

\bibitem[\protect\citeauthoryear{{Culhane}
  \textit{et~al.}}{2007}]{Culhaneetal2007SoPh}
\begin{barticle}
\bauthor{\bsnm{{Culhane}}, \binits{J.L.}},
\bauthor{\bsnm{{Harra}}, \binits{L.K.}},
\bauthor{\bsnm{{James}}, \binits{A.M.}},
\bauthor{\bsnm{{Al-Janabi}}, \binits{K.}},
\bauthor{\bsnm{{Bradley}}, \binits{L.J.}},
\bauthor{\bsnm{{Chaudry}}, \binits{R.A.}},
\bauthor{\bsnm{{Rees}}, \binits{K.}},
\bauthor{\bsnm{{Tandy}}, \binits{J.A.}},
\bauthor{\bsnm{{Thomas}}, \binits{P.}},
\bauthor{\bsnm{{Whillock}}, \binits{M.C.R.}},
\bauthor{\bsnm{{Winter}}, \binits{B.}},
\bauthor{\bsnm{{Doschek}}, \binits{G.A.}},
\bauthor{\bsnm{{Korendyke}}, \binits{C.M.}},
\bauthor{\bsnm{{Brown}}, \binits{C.M.}},
\bauthor{\bsnm{{Myers}}, \binits{S.}},
\bauthor{\bsnm{{Mariska}}, \binits{J.}},
\bauthor{\bsnm{{Seely}}, \binits{J.}},
\bauthor{\bsnm{{Lang}}, \binits{J.}},
\bauthor{\bsnm{{Kent}}, \binits{B.J.}},
\bauthor{\bsnm{{Shaughnessy}}, \binits{B.M.}},
\bauthor{\bsnm{{Young}}, \binits{P.R.}},
\bauthor{\bsnm{{Simnett}}, \binits{G.M.}},
\bauthor{\bsnm{{Castelli}}, \binits{C.M.}},
\bauthor{\bsnm{{Mahmoud}}, \binits{S.}},
\bauthor{\bsnm{{Mapson-Menard}}, \binits{H.}},
\bauthor{\bsnm{{Probyn}}, \binits{B.J.}},
\bauthor{\bsnm{{Thomas}}, \binits{R.J.}},
\bauthor{\bsnm{{Davila}}, \binits{J.}},
\bauthor{\bsnm{{Dere}}, \binits{K.}},
\bauthor{\bsnm{{Windt}}, \binits{D.}},
\bauthor{\bsnm{{Shea}}, \binits{J.}},
\bauthor{\bsnm{{Hagood}}, \binits{R.}},
\bauthor{\bsnm{{Moye}}, \binits{R.}},
\bauthor{\bsnm{{Hara}}, \binits{H.}},
\bauthor{\bsnm{{Watanabe}}, \binits{T.}},
\bauthor{\bsnm{{Matsuzaki}}, \binits{K.}},
\bauthor{\bsnm{{Kosugi}}, \binits{T.}},
\bauthor{\bsnm{{Hansteen}}, \binits{V.}},
\bauthor{\bsnm{{Wikstol}}, \binits{{\O}.}}:
\byear{2007},
\batitle{{The EUV Imaging Spectrometer for Hinode}}.
\bjtitle{\solphys}
\bvolume{243},
\bfpage{19}\,--\,\blpage{61}.
doi:\doiurl{10.1007/s01007-007-0293-1}.
\end{barticle}
\endbibitem

\bibitem[\protect\citeauthoryear{{Debosscher}
  \textit{et~al.}}{2007}]{2007A&A...475.1159D}
\begin{barticle}
\bauthor{\bsnm{{Debosscher}}, \binits{J.}},
\bauthor{\bsnm{{Sarro}}, \binits{L.M.}},
\bauthor{\bsnm{{Aerts}}, \binits{C.}},
\bauthor{\bsnm{{Cuypers}}, \binits{J.}},
\bauthor{\bsnm{{Vandenbussche}}, \binits{B.}},
\bauthor{\bsnm{{Garrido}}, \binits{R.}},
\bauthor{\bsnm{{Solano}}, \binits{E.}}:
\byear{2007},
\batitle{{Automated supervised classification of variable stars. I.
  Methodology}}.
\bjtitle{\aap}
\bvolume{475},
\bfpage{1159}\,--\,\blpage{1183}.
doi:\doiurl{10.1051/0004-6361:20077638}.
\end{barticle}
\endbibitem

\bibitem[\protect\citeauthoryear{{Debosscher}
  \textit{et~al.}}{2009}]{2009A&A...506..519D}
\begin{barticle}
\bauthor{\bsnm{{Debosscher}}, \binits{J.}},
\bauthor{\bsnm{{Sarro}}, \binits{L.M.}},
\bauthor{\bsnm{{L{\'o}pez}}, \binits{M.}},
\bauthor{\bsnm{{Deleuil}}, \binits{M.}},
\bauthor{\bsnm{{Aerts}}, \binits{C.}},
\bauthor{\bsnm{{Auvergne}}, \binits{M.}},
\bauthor{\bsnm{{Baglin}}, \binits{A.}},
\bauthor{\bsnm{{Baudin}}, \binits{F.}},
\bauthor{\bsnm{{Chadid}}, \binits{M.}},
\bauthor{\bsnm{{Charpinet}}, \binits{S.}},
\bauthor{\bsnm{{Cuypers}}, \binits{J.}},
\bauthor{\bsnm{{De Ridder}}, \binits{J.}},
\bauthor{\bsnm{{Garrido}}, \binits{R.}},
\bauthor{\bsnm{{Hubert}}, \binits{A.M.}},
\bauthor{\bsnm{{Janot-Pacheco}}, \binits{E.}},
\bauthor{\bsnm{{Jorda}}, \binits{L.}},
\bauthor{\bsnm{{Kaiser}}, \binits{A.}},
\bauthor{\bsnm{{Kallinger}}, \binits{T.}},
\bauthor{\bsnm{{Kollath}}, \binits{Z.}},
\bauthor{\bsnm{{Maceroni}}, \binits{C.}},
\bauthor{\bsnm{{Mathias}}, \binits{P.}},
\bauthor{\bsnm{{Michel}}, \binits{E.}},
\bauthor{\bsnm{{Moutou}}, \binits{C.}},
\bauthor{\bsnm{{Neiner}}, \binits{C.}},
\bauthor{\bsnm{{Ollivier}}, \binits{M.}},
\bauthor{\bsnm{{Samadi}}, \binits{R.}},
\bauthor{\bsnm{{Solano}}, \binits{E.}},
\bauthor{\bsnm{{Surace}}, \binits{C.}},
\bauthor{\bsnm{{Vandenbussche}}, \binits{B.}},
\bauthor{\bsnm{{Weiss}}, \binits{W.W.}}:
\byear{2009},
\batitle{{Automated supervised classification of variable stars in the CoRoT
  programme. Method and application to the first four exoplanet fields}}.
\bjtitle{\aap}
\bvolume{506},
\bfpage{519}\,--\,\blpage{534}.
doi:\doiurl{10.1051/0004-6361/200911618}.
\end{barticle}
\endbibitem

\bibitem[\protect\citeauthoryear{{Del Zanna}}{2012}]{DelZanna2012arXiv}
\begin{botherref}
\oauthor{\bsnm{{Del Zanna}}, \binits{G.}}:
2012,
{In-flight calibration of the Hinode EIS. Preliminary results}.
\textit{ArXiv e-prints}.
\end{botherref}
\endbibitem

\bibitem[\protect\citeauthoryear{{Delaboudini{\`e}re}
  \textit{et~al.}}{1995}]{Delaboudiniereetal1995SoPh}
\begin{barticle}
\bauthor{\bsnm{{Delaboudini{\`e}re}}, \binits{J.-P.}},
\bauthor{\bsnm{{Artzner}}, \binits{G.E.}},
\bauthor{\bsnm{{Brunaud}}, \binits{J.}},
\bauthor{\bsnm{{Gabriel}}, \binits{A.H.}},
\bauthor{\bsnm{{Hochedez}}, \binits{J.F.}},
\bauthor{\bsnm{{Millier}}, \binits{F.}},
\bauthor{\bsnm{{Song}}, \binits{X.Y.}},
\bauthor{\bsnm{{Au}}, \binits{B.}},
\bauthor{\bsnm{{Dere}}, \binits{K.P.}},
\bauthor{\bsnm{{Howard}}, \binits{R.A.}},
\bauthor{\bsnm{{Kreplin}}, \binits{R.}},
\bauthor{\bsnm{{Michels}}, \binits{D.J.}},
\bauthor{\bsnm{{Moses}}, \binits{J.D.}},
\bauthor{\bsnm{{Defise}}, \binits{J.M.}},
\bauthor{\bsnm{{Jamar}}, \binits{C.}},
\bauthor{\bsnm{{Rochus}}, \binits{P.}},
\bauthor{\bsnm{{Chauvineau}}, \binits{J.P.}},
\bauthor{\bsnm{{Marioge}}, \binits{J.P.}},
\bauthor{\bsnm{{Catura}}, \binits{R.C.}},
\bauthor{\bsnm{{Lemen}}, \binits{J.R.}},
\bauthor{\bsnm{{Shing}}, \binits{L.}},
\bauthor{\bsnm{{Stern}}, \binits{R.A.}},
\bauthor{\bsnm{{Gurman}}, \binits{J.B.}},
\bauthor{\bsnm{{Neupert}}, \binits{W.M.}},
\bauthor{\bsnm{{Maucherat}}, \binits{A.}},
\bauthor{\bsnm{{Clette}}, \binits{F.}},
\bauthor{\bsnm{{Cugnon}}, \binits{P.}},
\bauthor{\bsnm{{van Dessel}}, \binits{E.L.}}:
\byear{1995},
\batitle{{EIT: Extreme-Ultraviolet Imaging Telescope for the SOHO Mission}}.
\bjtitle{\solphys}
\bvolume{162},
\bfpage{291}\,--\,\blpage{312}.
doi:\doiurl{10.1007/BF00733432}.
\end{barticle}
\endbibitem

\bibitem[\protect\citeauthoryear{{Dere} \textit{et~al.}}{1997}]{Dereetal1997}
\begin{barticle}
\bauthor{\bsnm{{Dere}}, \binits{K.P.}},
\bauthor{\bsnm{{Landi}}, \binits{E.}},
\bauthor{\bsnm{{Mason}}, \binits{H.E.}},
\bauthor{\bsnm{{Monsignori Fossi}}, \binits{B.C.}},
\bauthor{\bsnm{{Young}}, \binits{P.R.}}:
\byear{1997},
\batitle{{CHIANTI - An atomic database for emission lines}}.
\bjtitle{\aaps}
\bvolume{125},
\bfpage{149}\,--\,\blpage{173}.
doi:\doiurl{10.1051/aas:1997368}.
\end{barticle}
\endbibitem

\bibitem[\protect\citeauthoryear{{Dere} \textit{et~al.}}{2009}]{Dereetal2009}
\begin{barticle}
\bauthor{\bsnm{{Dere}}, \binits{K.P.}},
\bauthor{\bsnm{{Landi}}, \binits{E.}},
\bauthor{\bsnm{{Young}}, \binits{P.R.}},
\bauthor{\bsnm{{Del Zanna}}, \binits{G.}},
\bauthor{\bsnm{{Landini}}, \binits{M.}},
\bauthor{\bsnm{{Mason}}, \binits{H.E.}}:
\byear{2009},
\batitle{{CHIANTI - An atomic database for emission lines. IX. Ionization
  rates, recombination rates, ionization equilibria for the elements hydrogen
  through zinc and updated atomic data}}.
\bjtitle{\aap}
\bvolume{498},
\bfpage{915}\,--\,\blpage{929}.
doi:\doiurl{10.1051/0004-6361/200911712}.
\end{barticle}
\endbibitem

\bibitem[\protect\citeauthoryear{{Doschek}
  \textit{et~al.}}{2010}]{Doscheketal2010ApJ}
\begin{barticle}
\bauthor{\bsnm{{Doschek}}, \binits{G.A.}},
\bauthor{\bsnm{{Landi}}, \binits{E.}},
\bauthor{\bsnm{{Warren}}, \binits{H.P.}},
\bauthor{\bsnm{{Harra}}, \binits{L.K.}}:
\byear{2010},
\batitle{{Bright points and jets in polar coronal holes observed by the
  Extreme-Ultraviolet Imaging Spectrometer on Hinode}}.
\bjtitle{\apj}
\bvolume{710},
\bfpage{1806}\,--\,\blpage{1824}.
doi:\doiurl{10.1088/0004-637X/710/2/1806}.
\end{barticle}
\endbibitem

\bibitem[\protect\citeauthoryear{{Dowdy}}{1993}]{1993ApJ...411..406D}
\begin{barticle}
\bauthor{\bsnm{{Dowdy}}, \binits{J.F.} \bsuffix{Jr.}}:
\byear{1993},
\batitle{{Observational evidence for hotter transition region loops within the
  supergranular network}}.
\bjtitle{\apj}
\bvolume{411},
\bfpage{406}\,--\,\blpage{409}.
doi:\doiurl{10.1086/172842}.
\end{barticle}
\endbibitem

\bibitem[\protect\citeauthoryear{{Doyle}, {Roussev}, and
  {Madjarska}}{2004}]{2004A&A...418L...9D}
\begin{barticle}
\bauthor{\bsnm{{Doyle}}, \binits{J.G.}},
\bauthor{\bsnm{{Roussev}}, \binits{I.I.}},
\bauthor{\bsnm{{Madjarska}}, \binits{M.S.}}:
\byear{2004},
\batitle{{New insight into the blinker phenomenon and the dynamics of the solar
  transition region}}.
\bjtitle{\aap}
\bvolume{418},
\bfpage{L9}\,--\,\blpage{L12}.
doi:\doiurl{10.1051/0004-6361:20040104}.
\end{barticle}
\endbibitem

\bibitem[\protect\citeauthoryear{{Doyle}
  \textit{et~al.}}{1998}]{Doyleetal1998SoPh}
\begin{barticle}
\bauthor{\bsnm{{Doyle}}, \binits{J.G.}},
\bauthor{\bsnm{{van den Oord}}, \binits{G.H.J.}},
\bauthor{\bsnm{{O'Shea}}, \binits{E.}},
\bauthor{\bsnm{{Banerjee}}, \binits{D.}}:
\byear{1998},
\batitle{{Waves in the solar transition region}}.
\bjtitle{\solphys}
\bvolume{181},
\bfpage{51}\,--\,\blpage{71}.
doi:\doiurl{10.1023/A:1005078018623.}.
\end{barticle}
\endbibitem

\bibitem[\protect\citeauthoryear{{Egamberdiev} and
  {Iakovkin}}{1983}]{EgamberdievIakovkin1983DoUzb}
\begin{barticle}
\bauthor{\bsnm{{Egamberdiev}}, \binits{S.A.}},
\bauthor{\bsnm{{Iakovkin}}, \binits{N.A.}}:
\byear{1983},
\batitle{{The luminiscence of bright X-ray points on the sun}}.
\bjtitle{{\it Dokl. Akad. Nauk Uzbek. SSR}}
\bvolume{7},
\bfpage{31}\,--\,\blpage{32}.
\end{barticle}
\endbibitem

\bibitem[\protect\citeauthoryear{{Gebbie}
  \textit{et~al.}}{1981}]{Gebbieetal1981ApJ}
\begin{barticle}
\bauthor{\bsnm{{Gebbie}}, \binits{K.B.}},
\bauthor{\bsnm{{Hill}}, \binits{F.}},
\bauthor{\bsnm{{November}}, \binits{L.J.}},
\bauthor{\bsnm{{Gurman}}, \binits{J.B.}},
\bauthor{\bsnm{{Shine}}, \binits{R.A.}},
\bauthor{\bsnm{{Woodgate}}, \binits{B.E.}},
\bauthor{\bsnm{{Athay}}, \binits{R.G.}},
\bauthor{\bsnm{{Tandberg-Hanssen}}, \binits{E.A.}},
\bauthor{\bsnm{{Toomre}}, \binits{J.}},
\bauthor{\bsnm{{Simon}}, \binits{G.W.}}:
\byear{1981},
\batitle{{Steady flows in the solar transition region observed with SMM}}.
\bjtitle{\apjl}
\bvolume{251},
\bfpage{L115}\,--\,\blpage{L118}.
doi:\doiurl{10.1086/183705}.
\end{barticle}
\endbibitem

\bibitem[\protect\citeauthoryear{{Golub}, {Krieger}, and
  {Vaiana}}{1976}]{1976SoPh...50..311G}
\begin{barticle}
\bauthor{\bsnm{{Golub}}, \binits{L.}},
\bauthor{\bsnm{{Krieger}}, \binits{A.S.}},
\bauthor{\bsnm{{Vaiana}}, \binits{G.S.}}:
\byear{1976},
\batitle{{Observation of spatial and temporal variations in X-ray bright point
  emergence patterns}}.
\bjtitle{\solphys}
\bvolume{50},
\bfpage{311}\,--\,\blpage{327}.
doi:\doiurl{10.1007/BF00155294}.
\end{barticle}
\endbibitem

\bibitem[\protect\citeauthoryear{{Golub} \textit{et~al.}}{1974}]{Golub1974}
\begin{barticle}
\bauthor{\bsnm{{Golub}}, \binits{L.}},
\bauthor{\bsnm{{Krieger}}, \binits{A.S.}},
\bauthor{\bsnm{{Silk}}, \binits{J.K.}},
\bauthor{\bsnm{{Timothy}}, \binits{A.F.}},
\bauthor{\bsnm{{Vaiana}}, \binits{G.S.}}:
\byear{1974},
\batitle{{Solar X-ray bright points}}.
\bjtitle{\apjl}
\bvolume{189},
\bfpage{L93}\,--\,\blpage{L97}.
doi:\doiurl{10.1086/181472}.
\end{barticle}
\endbibitem

\bibitem[\protect\citeauthoryear{{Grevesse}, {Asplund}, and
  {Sauval}}{2007}]{2007SSRv..130..105G}
\begin{barticle}
\bauthor{\bsnm{{Grevesse}}, \binits{N.}},
\bauthor{\bsnm{{Asplund}}, \binits{M.}},
\bauthor{\bsnm{{Sauval}}, \binits{A.J.}}:
\byear{2007},
\batitle{{The solar chemical composition}}.
\bjtitle{\ssr}
\bvolume{130},
\bfpage{105}\,--\,\blpage{114}.
doi:\doiurl{10.1007/s11214-007-9173-7}.
\end{barticle}
\endbibitem

\bibitem[\protect\citeauthoryear{{Habbal} and {Withbroe}}{1981}]{1981SoPh6977H}
\begin{barticle}
\bauthor{\bsnm{{Habbal}}, \binits{S.R.}},
\bauthor{\bsnm{{Withbroe}}, \binits{G.L.}}:
\byear{1981},
\batitle{{Spatial and temporal variations of EUV coronal bright points}}.
\bjtitle{\solphys}
\bvolume{69},
\bfpage{77}\,--\,\blpage{97}.
doi:\doiurl{10.1007/BF00151257}.
\end{barticle}
\endbibitem

\bibitem[\protect\citeauthoryear{{Habbal}, {Withbroe}, and
  {Dowdy}}{1990}]{Habbaletal1990ApJ}
\begin{barticle}
\bauthor{\bsnm{{Habbal}}, \binits{S.R.}},
\bauthor{\bsnm{{Withbroe}}, \binits{G.L.}},
\bauthor{\bsnm{{Dowdy}}, \binits{J.F.} \bsuffix{Jr.}}:
\byear{1990},
\batitle{{A comparison between bright points in a coronal hole and a quiet-sun
  region}}.
\bjtitle{\apj}
\bvolume{352},
\bfpage{333}\,--\,\blpage{342}.
doi:\doiurl{10.1086/168540}.
\end{barticle}
\endbibitem

\bibitem[\protect\citeauthoryear{{Harrison}}{1997}]{1997SoPh..175..467H}
\begin{barticle}
\bauthor{\bsnm{{Harrison}}, \binits{R.A.}}:
\byear{1997},
\batitle{{EUV Blinkers: The significance of variations in the extreme
  ultraviolet quiet Sun}}.
\bjtitle{\solphys}
\bvolume{175},
\bfpage{467}\,--\,\blpage{485}.
doi:\doiurl{10.1023/A:1004964707047}.
\end{barticle}
\endbibitem

\bibitem[\protect\citeauthoryear{{Harrison}
  \textit{et~al.}}{1997}]{1997AdSpR..20.2239H}
\begin{barticle}
\bauthor{\bsnm{{Harrison}}, \binits{R.A.}},
\bauthor{\bsnm{{Fludra}}, \binits{A.}},
\bauthor{\bsnm{{Sawyer}}, \binits{E.C.}},
\bauthor{\bsnm{{Culhane}}, \binits{J.L.}},
\bauthor{\bsnm{{Norman}}, \binits{K.}},
\bauthor{\bsnm{{Poland}}, \binits{A.I.}},
\bauthor{\bsnm{{Thompson}}, \binits{W.T.}},
\bauthor{\bsnm{{Kjeldseth-Moe}}, \binits{O.}},
\bauthor{\bsnm{{Aschenbach}}, \binits{B.}},
\bauthor{\bsnm{{Huber}}, \binits{M.C.E.}},
\bauthor{\bsnm{{Gabriel}}, \binits{A.H.}},
\bauthor{\bsnm{{Mason}}, \binits{H.E.}}:
\byear{1997},
\batitle{{Extreme ultraviolet observations of the solar corona: First results
  from the coronal diagnostic spectrometer on SOHO}}.
\bjtitle{Adv. Space Res.}
\bvolume{20},
\bfpage{2239}\,--\,\blpage{2248}.
doi:\doiurl{10.1016/S0273-1177(97)01059-4}.
\end{barticle}
\endbibitem

\bibitem[\protect\citeauthoryear{{Harrison}
  \textit{et~al.}}{1999}]{1999A&A...351.1115H}
\begin{barticle}
\bauthor{\bsnm{{Harrison}}, \binits{R.A.}},
\bauthor{\bsnm{{Lang}}, \binits{J.}},
\bauthor{\bsnm{{Brooks}}, \binits{D.H.}},
\bauthor{\bsnm{{Innes}}, \binits{D.E.}}:
\byear{1999},
\batitle{{A study of extreme ultraviolet blinker activity}}.
\bjtitle{\aap}
\bvolume{351},
\bfpage{1115}\,--\,\blpage{1132}.
\end{barticle}
\endbibitem

\bibitem[\protect\citeauthoryear{{Harrison}
  \textit{et~al.}}{2003}]{2003A&A...409..755H}
\begin{barticle}
\bauthor{\bsnm{{Harrison}}, \binits{R.A.}},
\bauthor{\bsnm{{Harra}}, \binits{L.K.}},
\bauthor{\bsnm{{Brkovi{\'c}}}, \binits{A.}},
\bauthor{\bsnm{{Parnell}}, \binits{C.E.}}:
\byear{2003},
\batitle{{A study of the unification of quiet-Sun transient-event phenomena}}.
\bjtitle{\aap}
\bvolume{409},
\bfpage{755}\,--\,\blpage{764}.
doi:\doiurl{10.1051/0004-6361:20031072}.
\end{barticle}
\endbibitem

\bibitem[\protect\citeauthoryear{{Heggland}, {De Pontieu}, and
  {Hansteen}}{2009}]{Hegglandetal2009ApJ}
\begin{barticle}
\bauthor{\bsnm{{Heggland}}, \binits{L.}},
\bauthor{\bsnm{{De Pontieu}}, \binits{B.}},
\bauthor{\bsnm{{Hansteen}}, \binits{V.H.}}:
\byear{2009},
\batitle{{Observational signatures of simulated reconnection events in the
  solar chromosphere and transition region}}.
\bjtitle{\apj}
\bvolume{702},
\bfpage{1}\,--\,\blpage{18}.
doi:\doiurl{10.1088/0004-637X/702/1/1}.
\end{barticle}
\endbibitem

\bibitem[\protect\citeauthoryear{{Kamio}
  \textit{et~al.}}{2010}]{2010SoPh..266..209K}
\begin{barticle}
\bauthor{\bsnm{{Kamio}}, \binits{S.}},
\bauthor{\bsnm{{Hara}}, \binits{H.}},
\bauthor{\bsnm{{Watanabe}}, \binits{T.}},
\bauthor{\bsnm{{Fredvik}}, \binits{T.}},
\bauthor{\bsnm{{Hansteen}}, \binits{V.H.}}:
\byear{2010},
\batitle{{Modeling of EIS spectrum drift from instrumental temperatures}}.
\bjtitle{\solphys}
\bvolume{266},
\bfpage{209}\,--\,\blpage{223}.
doi:\doiurl{10.1007/s11207-010-9603-7}.
\end{barticle}
\endbibitem

\bibitem[\protect\citeauthoryear{{Kamio}
  \textit{et~al.}}{2011}]{2011A&A...529A..21K}
\begin{barticle}
\bauthor{\bsnm{{Kamio}}, \binits{S.}},
\bauthor{\bsnm{{Curdt}}, \binits{W.}},
\bauthor{\bsnm{{Teriaca}}, \binits{L.}},
\bauthor{\bsnm{{Innes}}, \binits{D.E.}}:
\byear{2011},
\batitle{{Evolution of microflares associated with bright points in coronal
  holes and in quiet regions}}.
\bjtitle{\aap}
\bvolume{529},
\bfpage{A21}.
doi:\doiurl{10.1051/0004-6361/201015715}.
\end{barticle}
\endbibitem

\bibitem[\protect\citeauthoryear{{Kankelborg}, {Walker}, and
  {Hoover}}{1997}]{1997ApJ...491..952K}
\begin{barticle}
\bauthor{\bsnm{{Kankelborg}}, \binits{C.C.}},
\bauthor{\bsnm{{Walker}}, \binits{A.B.C.} \bsuffix{II}},
\bauthor{\bsnm{{Hoover}}, \binits{R.B.}}:
\byear{1997},
\batitle{{Observation and modeling of soft X-ray bright points. II.
  Determination of temperature and energy balance}}.
\bjtitle{\apj}
\bvolume{491},
\bfpage{952}\,--\,\blpage{966}.
doi:\doiurl{10.1086/304976}.
\end{barticle}
\endbibitem

\bibitem[\protect\citeauthoryear{{Kankelborg}
  \textit{et~al.}}{1996}]{1996ApJ...466..529K}
\begin{barticle}
\bauthor{\bsnm{{Kankelborg}}, \binits{C.C.}},
\bauthor{\bsnm{{Walker}}, \binits{A.B.C.} \bsuffix{II}},
\bauthor{\bsnm{{Hoover}}, \binits{R.B.}},
\bauthor{\bsnm{{Barbee}}, \binits{T.W.} \bsuffix{Jr.}}:
\byear{1996},
\batitle{{Observation and modeling of soft X-ray bright points. I. Initial
  results}}.
\bjtitle{\apj}
\bvolume{466},
\bfpage{529}.
doi:\doiurl{10.1086/177529}.
\end{barticle}
\endbibitem

\bibitem[\protect\citeauthoryear{{Koutchmy}
  \textit{et~al.}}{1997}]{Koutchmyetal1997A&A}
\begin{barticle}
\bauthor{\bsnm{{Koutchmy}}, \binits{S.}},
\bauthor{\bsnm{{Hara}}, \binits{H.}},
\bauthor{\bsnm{{Suematsu}}, \binits{Y.}},
\bauthor{\bsnm{{Reardon}}, \binits{K.}}:
\byear{1997},
\batitle{{SXR Coronal flashes.}}
\bjtitle{\aap}
\bvolume{320},
\bfpage{L33}\,--\,\blpage{L36}.
\end{barticle}
\endbibitem

\bibitem[\protect\citeauthoryear{{Krieger}, {Vaiana}, and {van
  Speybroeck}}{1971}]{Krieger1971}
\begin{bchapter}
\bauthor{\bsnm{{Krieger}}, \binits{A.S.}},
\bauthor{\bsnm{{Vaiana}}, \binits{G.S.}},
\bauthor{\bsnm{{van Speybroeck}}, \binits{L.P.}}:
\byear{1971},
\bctitle{{The X-ray corona and the photospheric magnetic field}}.
In: \beditor{\bsnm{{Howard}}, \binits{R.}} (ed.)
\bbtitle{Solar Magnetic Fields},
\bsertitle{IAU Symp.}
\bseriesno{43},
\bfpage{397}.
\end{bchapter}
\endbibitem

\bibitem[\protect\citeauthoryear{{Krucker}
  \textit{et~al.}}{1997}]{1997ApJ...488..499K}
\begin{barticle}
\bauthor{\bsnm{{Krucker}}, \binits{S.}},
\bauthor{\bsnm{{Benz}}, \binits{A.O.}},
\bauthor{\bsnm{{Bastian}}, \binits{T.S.}},
\bauthor{\bsnm{{Acton}}, \binits{L.W.}}:
\byear{1997},
\batitle{{X-Ray network flares of the quiet Sun}}.
\bjtitle{\apj}
\bvolume{488},
\bfpage{499}\,--\,\blpage{505}.
doi:\doiurl{10.1086/304686}.
\end{barticle}
\endbibitem

\bibitem[\protect\citeauthoryear{{Kwon}
  \textit{et~al.}}{2012}]{Kwonetal2012ApJ}
\begin{barticle}
\bauthor{\bsnm{{Kwon}}, \binits{R.-Y.}},
\bauthor{\bsnm{{Chae}}, \binits{J.}},
\bauthor{\bsnm{{Davila}}, \binits{J.M.}},
\bauthor{\bsnm{{Zhang}}, \binits{J.}},
\bauthor{\bsnm{{Moon}}, \binits{Y.-J.}},
\bauthor{\bsnm{{Poomvises}}, \binits{W.}},
\bauthor{\bsnm{{Jones}}, \binits{S.I.}}:
\byear{2012},
\batitle{{Three-dimensional structure and evolution of extreme-ultraviolet
  bright points observed by STEREO/SECCHI/EUVI}}.
\bjtitle{\apj}
\bvolume{757},
\bfpage{167}.
doi:\doiurl{10.1088/0004-637X/757/2/167}.
\end{barticle}
\endbibitem

\bibitem[\protect\citeauthoryear{{Lee} \textit{et~al.}}{2011}]{Leeetal2011ApJ}
\begin{barticle}
\bauthor{\bsnm{{Lee}}, \binits{K.-S.}},
\bauthor{\bsnm{{Moon}}, \binits{Y.-J.}},
\bauthor{\bsnm{{Kim}}, \binits{S.}},
\bauthor{\bsnm{{Choe}}, \binits{G.S.}},
\bauthor{\bsnm{{Cho}}, \binits{K.-S.}},
\bauthor{\bsnm{{Imada}}, \binits{S.}}:
\byear{2011},
\batitle{{Two types of extreme-ultraviolet brightenings in AR 10926 observed by
  Hinode/EIS}}.
\bjtitle{\apj}
\bvolume{736},
\bfpage{15}.
doi:\doiurl{10.1088/0004-637X/736/1/15}.
\end{barticle}
\endbibitem

\bibitem[\protect\citeauthoryear{{Long}
  \textit{et~al.}}{2012}]{2012PASP..124..280L}
\begin{barticle}
\bauthor{\bsnm{{Long}}, \binits{J.P.}},
\bauthor{\bsnm{{Karoui}}, \binits{N.E.}},
\bauthor{\bsnm{{Rice}}, \binits{J.A.}},
\bauthor{\bsnm{{Richards}}, \binits{J.W.}},
\bauthor{\bsnm{{Bloom}}, \binits{J.S.}}:
\byear{2012},
\batitle{{Optimizing automated classification of variable stars in new synoptic
  surveys}}.
\bjtitle{\pasp}
\bvolume{124},
\bfpage{280}\,--\,\blpage{295}.
doi:\doiurl{10.1086/664960}.
\end{barticle}
\endbibitem

\bibitem[\protect\citeauthoryear{{Longcope}}{1998}]{1998ApJ...507..433L}
\begin{barticle}
\bauthor{\bsnm{{Longcope}}, \binits{D.W.}}:
\byear{1998},
\batitle{{A model for current sheets and reconnection in X-Ray bright points}}.
\bjtitle{\apj}
\bvolume{507},
\bfpage{433}\,--\,\blpage{442}.
doi:\doiurl{10.1086/306319}.
\end{barticle}
\endbibitem

\bibitem[\protect\citeauthoryear{{Longcope} and
  {Kankelborg}}{1999}]{LongcopeKankelborg1999ApJ}
\begin{barticle}
\bauthor{\bsnm{{Longcope}}, \binits{D.W.}},
\bauthor{\bsnm{{Kankelborg}}, \binits{C.C.}}:
\byear{1999},
\batitle{{Coronal heating by collision and cancellation of magnetic elements}}.
\bjtitle{\apj}
\bvolume{524},
\bfpage{483}\,--\,\blpage{495}.
doi:\doiurl{10.1086/307792}.
\end{barticle}
\endbibitem

\bibitem[\protect\citeauthoryear{{Longcope}
  \textit{et~al.}}{2001}]{2001ApJ...553..429L}
\begin{barticle}
\bauthor{\bsnm{{Longcope}}, \binits{D.W.}},
\bauthor{\bsnm{{Kankelborg}}, \binits{C.C.}},
\bauthor{\bsnm{{Nelson}}, \binits{J.L.}},
\bauthor{\bsnm{{Pevtsov}}, \binits{A.A.}}:
\byear{2001},
\batitle{{Evidence of separator reconnection in a survey of X-ray bright
  points}}.
\bjtitle{\apj}
\bvolume{553},
\bfpage{429}\,--\,\blpage{439}.
doi:\doiurl{10.1086/320667}.
\end{barticle}
\endbibitem

\bibitem[\protect\citeauthoryear{{Madjarska} and
  {Doyle}}{2003}]{2003A&A...403..731M}
\begin{barticle}
\bauthor{\bsnm{{Madjarska}}, \binits{M.S.}},
\bauthor{\bsnm{{Doyle}}, \binits{J.G.}}:
\byear{2003},
\batitle{{Simultaneous observations of solar transition region blinkers and
  explosive events by SUMER, CDS and BBSO. Are blinkers, explosive events and
  spicules the same phenomenon?}}
\bjtitle{\aap}
\bvolume{403},
\bfpage{731}\,--\,\blpage{741}.
doi:\doiurl{10.1051/0004-6361:20030397}.
\end{barticle}
\endbibitem

\bibitem[\protect\citeauthoryear{{Madjarska}, {Doyle}, and {De
  Pontieu}}{2009}]{2009ApJ...701..253M}
\begin{barticle}
\bauthor{\bsnm{{Madjarska}}, \binits{M.S.}},
\bauthor{\bsnm{{Doyle}}, \binits{J.G.}},
\bauthor{\bsnm{{De Pontieu}}, \binits{B.}}:
\byear{2009},
\batitle{{Explosive events associated with a surge}}.
\bjtitle{\apj}
\bvolume{701},
\bfpage{253}\,--\,\blpage{259}.
doi:\doiurl{10.1088/0004-637X/701/1/253}.
\end{barticle}
\endbibitem

\bibitem[\protect\citeauthoryear{{Madjarska}
  \textit{et~al.}}{2003}]{Madjarskaetal2003A&A}
\begin{barticle}
\bauthor{\bsnm{{Madjarska}}, \binits{M.S.}},
\bauthor{\bsnm{{Doyle}}, \binits{J.G.}},
\bauthor{\bsnm{{Teriaca}}, \binits{L.}},
\bauthor{\bsnm{{Banerjee}}, \binits{D.}}:
\byear{2003},
\batitle{{An EUV bright point as seen by SUMER, CDS, MDI and EIT on-board
  SoHO}}.
\bjtitle{\aap}
\bvolume{398},
\bfpage{775}\,--\,\blpage{784}.
doi:\doiurl{10.1051/0004-6361:20021732}.
\end{barticle}
\endbibitem

\bibitem[\protect\citeauthoryear{{Madjarska}
  \textit{et~al.}}{2012}]{Madjarskaetal2012A&A}
\begin{barticle}
\bauthor{\bsnm{{Madjarska}}, \binits{M.S.}},
\bauthor{\bsnm{{Huang}}, \binits{Z.}},
\bauthor{\bsnm{{Doyle}}, \binits{J.G.}},
\bauthor{\bsnm{{Subramanian}}, \binits{S.}}:
\byear{2012},
\batitle{{Coronal hole boundaries evolution at small scales. III. EIS and SUMER
  views}}.
\bjtitle{\aap}
\bvolume{545},
\bfpage{A67}.
doi:\doiurl{10.1051/0004-6361/201219516}.
\end{barticle}
\endbibitem

\bibitem[\protect\citeauthoryear{{Mariska}
  \textit{et~al.}}{2007}]{Mariskaetal2007PASJ}
\begin{barticle}
\bauthor{\bsnm{{Mariska}}, \binits{J.T.}},
\bauthor{\bsnm{{Warren}}, \binits{H.P.}},
\bauthor{\bsnm{{Ugarte-Urra}}, \binits{I.}},
\bauthor{\bsnm{{Brooks}}, \binits{D.H.}},
\bauthor{\bsnm{{Williams}}, \binits{D.R.}},
\bauthor{\bsnm{{Hara}}, \binits{H.}}:
\byear{2007},
\batitle{{Hinode EUV Imaging Spectrometer observations of solar active region
  dynamics}}.
\bjtitle{\pasj}
\bvolume{59},
\bfpage{S713}\,--\,\blpage{S719}.
\end{barticle}
\endbibitem

\bibitem[\protect\citeauthoryear{{Matijevi{\v c}}
  \textit{et~al.}}{2012}]{2012AJ....143..123M}
\begin{barticle}
\bauthor{\bsnm{{Matijevi{\v c}}}, \binits{G.}},
\bauthor{\bsnm{{Pr{\v s}a}}, \binits{A.}},
\bauthor{\bsnm{{Orosz}}, \binits{J.A.}},
\bauthor{\bsnm{{Welsh}}, \binits{W.F.}},
\bauthor{\bsnm{{Bloemen}}, \binits{S.}},
\bauthor{\bsnm{{Barclay}}, \binits{T.}}:
\byear{2012},
\batitle{{Kepler eclipsing binary stars. III. Classification of Kepler
  eclipsing binary light curves with locally linear embedding}}.
\bjtitle{\aj}
\bvolume{143},
\bfpage{123}.
doi:\doiurl{10.1088/0004-6256/143/5/123}.
\end{barticle}
\endbibitem

\bibitem[\protect\citeauthoryear{{McClymont} and
  {Craig}}{1987}]{McClymontCraig1987}
\begin{barticle}
\bauthor{\bsnm{{McClymont}}, \binits{A.N.}},
\bauthor{\bsnm{{Craig}}, \binits{I.J.D.}}:
\byear{1987},
\batitle{{Fast downflows in the solar transition region explained}}.
\bjtitle{\apj}
\bvolume{312},
\bfpage{402}\,--\,\blpage{411}.
doi:\doiurl{10.1086/164885}.
\end{barticle}
\endbibitem

\bibitem[\protect\citeauthoryear{{McIntosh} and
  {Gurman}}{2005}]{2005SoPh..228..285M}
\begin{barticle}
\bauthor{\bsnm{{McIntosh}}, \binits{S.W.}},
\bauthor{\bsnm{{Gurman}}, \binits{J.B.}}:
\byear{2005},
\batitle{{Nine years of EUV bright points}}.
\bjtitle{\solphys}
\bvolume{228},
\bfpage{285}\,--\,\blpage{299}.
doi:\doiurl{10.1007/s11207-005-4725-z}.
\end{barticle}
\endbibitem

\bibitem[\protect\citeauthoryear{{Morgan}
  \textit{et~al.}}{2012}]{2012ApJ...746..170M}
\begin{barticle}
\bauthor{\bsnm{{Morgan}}, \binits{A.N.}},
\bauthor{\bsnm{{Long}}, \binits{J.}},
\bauthor{\bsnm{{Richards}}, \binits{J.W.}},
\bauthor{\bsnm{{Broderick}}, \binits{T.}},
\bauthor{\bsnm{{Butler}}, \binits{N.R.}},
\bauthor{\bsnm{{Bloom}}, \binits{J.S.}}:
\byear{2012},
\batitle{{Rapid, machine-learned resource allocation: Application to
  high-redshift gamma-ray burst follow-up}}.
\bjtitle{\apj}
\bvolume{746},
\bfpage{170}.
doi:\doiurl{10.1088/0004-637X/746/2/170}.
\end{barticle}
\endbibitem

\bibitem[\protect\citeauthoryear{{O'Dwyer}
  \textit{et~al.}}{2010}]{2010A&A...521A..21O}
\begin{barticle}
\bauthor{\bsnm{{O'Dwyer}}, \binits{B.}},
\bauthor{\bsnm{{Del Zanna}}, \binits{G.}},
\bauthor{\bsnm{{Mason}}, \binits{H.E.}},
\bauthor{\bsnm{{Weber}}, \binits{M.A.}},
\bauthor{\bsnm{{Tripathi}}, \binits{D.}}:
\byear{2010},
\batitle{{SDO/AIA response to coronal hole, quiet Sun, active region, and flare
  plasma}}.
\bjtitle{\aap}
\bvolume{521},
\bfpage{A21}.
doi:\doiurl{10.1051/0004-6361/201014872}.
\end{barticle}
\endbibitem

\bibitem[\protect\citeauthoryear{{Pariat}, {Antiochos}, and
  {DeVore}}{2010}]{Pariatetal2010ApJ}
\begin{barticle}
\bauthor{\bsnm{{Pariat}}, \binits{E.}},
\bauthor{\bsnm{{Antiochos}}, \binits{S.K.}},
\bauthor{\bsnm{{DeVore}}, \binits{C.R.}}:
\byear{2010},
\batitle{{Three-dimensional modeling of quasi-homologous solar jets}}.
\bjtitle{\apj}
\bvolume{714},
\bfpage{1762}\,--\,\blpage{1778}.
doi:\doiurl{10.1088/0004-637X/714/2/1762}.
\end{barticle}
\endbibitem

\bibitem[\protect\citeauthoryear{{Parnell}, {Bewsher}, and
  {Harrison}}{}]{2002SoPh..206..249P}
\begin{botherref}
\oauthor{\bsnm{{Parnell}}, \binits{C.E.}},
\oauthor{\bsnm{{Bewsher}}, \binits{D.}},
\oauthor{\bsnm{{Harrison}}, \binits{R.A.}}
doi:\doiurl{10.1023/A:1015094119974.}.
\end{botherref}
\endbibitem

\bibitem[\protect\citeauthoryear{{P{\'e}rez-Su{\'a}rez}
  \textit{et~al.}}{2008}]{Perez-Suarez2008A&A}
\begin{barticle}
\bauthor{\bsnm{{P{\'e}rez-Su{\'a}rez}}, \binits{D.}},
\bauthor{\bsnm{{Maclean}}, \binits{R.C.}},
\bauthor{\bsnm{{Doyle}}, \binits{J.G.}},
\bauthor{\bsnm{{Madjarska}}, \binits{M.S.}}:
\byear{2008},
\batitle{{The structure and dynamics of a bright point as seen with Hinode,
  SoHO and TRACE}}.
\bjtitle{\aap}
\bvolume{492},
\bfpage{575}\,--\,\blpage{583}.
doi:\doiurl{10.1051/0004-6361:200809507}.
\end{barticle}
\endbibitem

\bibitem[\protect\citeauthoryear{{Pevtsov}
  \textit{et~al.}}{2003}]{2003ApJ...598.1387P}
\begin{barticle}
\bauthor{\bsnm{{Pevtsov}}, \binits{A.A.}},
\bauthor{\bsnm{{Fisher}}, \binits{G.H.}},
\bauthor{\bsnm{{Acton}}, \binits{L.W.}},
\bauthor{\bsnm{{Longcope}}, \binits{D.W.}},
\bauthor{\bsnm{{Johns-Krull}}, \binits{C.M.}},
\bauthor{\bsnm{{Kankelborg}}, \binits{C.C.}},
\bauthor{\bsnm{{Metcalf}}, \binits{T.R.}}:
\byear{2003},
\batitle{{The relationship between X-ray radiance and magnetic flux}}.
\bjtitle{\apj}
\bvolume{598},
\bfpage{1387}\,--\,\blpage{1391}.
doi:\doiurl{10.1086/378944}.
\end{barticle}
\endbibitem

\bibitem[\protect\citeauthoryear{{Priest}, {Parnell}, and
  {Martin}}{1994}]{1994ApJ...427..459P}
\begin{barticle}
\bauthor{\bsnm{{Priest}}, \binits{E.R.}},
\bauthor{\bsnm{{Parnell}}, \binits{C.E.}},
\bauthor{\bsnm{{Martin}}, \binits{S.F.}}:
\byear{1994},
\batitle{{A converging flux model of an X-ray bright point and an associated
  canceling magnetic feature}}.
\bjtitle{\apj}
\bvolume{427},
\bfpage{459}\,--\,\blpage{474}.
doi:\doiurl{10.1086/174157}.
\end{barticle}
\endbibitem

\bibitem[\protect\citeauthoryear{{Raouafi}
  \textit{et~al.}}{2010}]{Raouafietal2010ApJ}
\begin{barticle}
\bauthor{\bsnm{{Raouafi}}, \binits{N.-E.}},
\bauthor{\bsnm{{Georgoulis}}, \binits{M.K.}},
\bauthor{\bsnm{{Rust}}, \binits{D.M.}},
\bauthor{\bsnm{{Bernasconi}}, \binits{P.N.}}:
\byear{2010},
\batitle{{Micro-sigmoids as progenitors of coronal jets: Is eruptive activity
  self-similarly multi-scaled?}}
\bjtitle{\apj}
\bvolume{718},
\bfpage{981}\,--\,\blpage{987}.
doi:\doiurl{10.1088/0004-637X/718/2/981}.
\end{barticle}
\endbibitem

\bibitem[\protect\citeauthoryear{{Richards}
  \textit{et~al.}}{2011}]{2011ApJ...733...10R}
\begin{barticle}
\bauthor{\bsnm{{Richards}}, \binits{J.W.}},
\bauthor{\bsnm{{Starr}}, \binits{D.L.}},
\bauthor{\bsnm{{Butler}}, \binits{N.R.}},
\bauthor{\bsnm{{Bloom}}, \binits{J.S.}},
\bauthor{\bsnm{{Brewer}}, \binits{J.M.}},
\bauthor{\bsnm{{Crellin-Quick}}, \binits{A.}},
\bauthor{\bsnm{{Higgins}}, \binits{J.}},
\bauthor{\bsnm{{Kennedy}}, \binits{R.}},
\bauthor{\bsnm{{Rischard}}, \binits{M.}}:
\byear{2011},
\batitle{{On machine-learned classification of variable stars with sparse and
  noisy time-series data}}.
\bjtitle{\apj}
\bvolume{733},
\bfpage{10}.
doi:\doiurl{10.1088/0004-637X/733/1/10}.
\end{barticle}
\endbibitem

\bibitem[\protect\citeauthoryear{{Richards}
  \textit{et~al.}}{2012}]{2012ApJ...744..192R}
\begin{barticle}
\bauthor{\bsnm{{Richards}}, \binits{J.W.}},
\bauthor{\bsnm{{Starr}}, \binits{D.L.}},
\bauthor{\bsnm{{Brink}}, \binits{H.}},
\bauthor{\bsnm{{Miller}}, \binits{A.A.}},
\bauthor{\bsnm{{Bloom}}, \binits{J.S.}},
\bauthor{\bsnm{{Butler}}, \binits{N.R.}},
\bauthor{\bsnm{{James}}, \binits{J.B.}},
\bauthor{\bsnm{{Long}}, \binits{J.P.}},
\bauthor{\bsnm{{Rice}}, \binits{J.}}:
\byear{2012},
\batitle{{Active learning to overcome sample selection bias: Application to
  photometric variable star classification}}.
\bjtitle{\apj}
\bvolume{744},
\bfpage{192}.
doi:\doiurl{10.1088/0004-637X/744/2/192}.
\end{barticle}
\endbibitem

\bibitem[\protect\citeauthoryear{{Romano}
  \textit{et~al.}}{2012}]{Romanoetal2012SoPh}
\begin{barticle}
\bauthor{\bsnm{{Romano}}, \binits{P.}},
\bauthor{\bsnm{{Berrilli}}, \binits{F.}},
\bauthor{\bsnm{{Criscuoli}}, \binits{S.}},
\bauthor{\bsnm{{Del Moro}}, \binits{D.}},
\bauthor{\bsnm{{Ermolli}}, \binits{I.}},
\bauthor{\bsnm{{Giorgi}}, \binits{F.}},
\bauthor{\bsnm{{Viticchi{\'e}}}, \binits{B.}},
\bauthor{\bsnm{{Zuccarello}}, \binits{F.}}:
\byear{2012},
\batitle{{A comparative analysis of photospheric bright points in an active
  region and in the quiet Sun}}.
\bjtitle{\solphys}
\bvolume{280},
\bfpage{407}\,--\,\blpage{416}.
doi:\doiurl{10.1007/s11207-012-9942-7}.
\end{barticle}
\endbibitem

\bibitem[\protect\citeauthoryear{{S{\'a}nchez Almeida}
  \textit{et~al.}}{2010}]{Sanchezetal2010ApJ}
\begin{barticle}
\bauthor{\bsnm{{S{\'a}nchez Almeida}}, \binits{J.}},
\bauthor{\bsnm{{Bonet}}, \binits{J.A.}},
\bauthor{\bsnm{{Viticchi{\'e}}}, \binits{B.}},
\bauthor{\bsnm{{Del Moro}}, \binits{D.}}:
\byear{2010},
\batitle{{Magnetic bright points in the quiet Sun}}.
\bjtitle{\apjl}
\bvolume{715},
\bfpage{L26}\,--\,\blpage{L29}.
doi:\doiurl{10.1088/2041-8205/715/1/L26}.
\end{barticle}
\endbibitem

\bibitem[\protect\citeauthoryear{{Sarro} and
  {Berihuete}}{2011}]{SarroBerihuete2011}
\begin{barticle}
\bauthor{\bsnm{{Sarro}}, \binits{L.M.}},
\bauthor{\bsnm{{Berihuete}}, \binits{A.}}:
\byear{2011},
\batitle{{Statistical techniques for the detection and analysis of solar
  explosive events}}.
\bjtitle{\aap}
\bvolume{528},
\bfpage{A62}.
doi:\doiurl{10.1051/0004-6361/201014894}.
\end{barticle}
\endbibitem

\bibitem[\protect\citeauthoryear{{Sarro}
  \textit{et~al.}}{2009a}]{2009A&A...494..739S}
\begin{barticle}
\bauthor{\bsnm{{Sarro}}, \binits{L.M.}},
\bauthor{\bsnm{{Debosscher}}, \binits{J.}},
\bauthor{\bsnm{{L{\'o}pez}}, \binits{M.}},
\bauthor{\bsnm{{Aerts}}, \binits{C.}}:
\byear{2009}a,
\batitle{{Automated supervised classification of variable stars. II.
  Application to the OGLE database}}.
\bjtitle{\aap}
\bvolume{494},
\bfpage{739}\,--\,\blpage{768}.
doi:\doiurl{10.1051/0004-6361:200809918}.
\end{barticle}
\endbibitem

\bibitem[\protect\citeauthoryear{{Sarro}
  \textit{et~al.}}{2009b}]{2009A&A...506..535S}
\begin{barticle}
\bauthor{\bsnm{{Sarro}}, \binits{L.M.}},
\bauthor{\bsnm{{Debosscher}}, \binits{J.}},
\bauthor{\bsnm{{Aerts}}, \binits{C.}},
\bauthor{\bsnm{{L{\'o}pez}}, \binits{M.}}:
\byear{2009}b,
\batitle{{Comparative clustering analysis of variable stars in the Hipparcos,
  OGLE Large Magellanic Cloud, and CoRoT exoplanet databases}}.
\bjtitle{\aap}
\bvolume{506},
\bfpage{535}\,--\,\blpage{568}.
doi:\doiurl{10.1051/0004-6361/200912009}.
\end{barticle}
\endbibitem

\bibitem[\protect\citeauthoryear{{Scherrer}
  \textit{et~al.}}{1995}]{Scherreretal1995SoPh}
\begin{barticle}
\bauthor{\bsnm{{Scherrer}}, \binits{P.H.}},
\bauthor{\bsnm{{Bogart}}, \binits{R.S.}},
\bauthor{\bsnm{{Bush}}, \binits{R.I.}},
\bauthor{\bsnm{{Hoeksema}}, \binits{J.T.}},
\bauthor{\bsnm{{Kosovichev}}, \binits{A.G.}},
\bauthor{\bsnm{{Schou}}, \binits{J.}},
\bauthor{\bsnm{{Rosenberg}}, \binits{W.}},
\bauthor{\bsnm{{Springer}}, \binits{L.}},
\bauthor{\bsnm{{Tarbell}}, \binits{T.D.}},
\bauthor{\bsnm{{Title}}, \binits{A.}},
\bauthor{\bsnm{{Wolfson}}, \binits{C.J.}},
\bauthor{\bsnm{{Zayer}}, \binits{I.}},
\bauthor{\bsnm{{MDI Engineering Team}}}:
\byear{1995},
\batitle{{The Solar Oscillations Investigation - Michelson Doppler Imager}}.
\bjtitle{\solphys}
\bvolume{162},
\bfpage{129}\,--\,\blpage{188}.
doi:\doiurl{10.1007/BF00733429}.
\end{barticle}
\endbibitem

\bibitem[\protect\citeauthoryear{{Shimizu}
  \textit{et~al.}}{2007}]{Shimizuetal2007}
\begin{barticle}
\bauthor{\bsnm{{Shimizu}}, \binits{T.}},
\bauthor{\bsnm{{Katsukawa}}, \binits{Y.}},
\bauthor{\bsnm{{Matsuzaki}}, \binits{K.}},
\bauthor{\bsnm{{Ichimoto}}, \binits{K.}},
\bauthor{\bsnm{{Kano}}, \binits{R.}},
\bauthor{\bsnm{{Deluca}}, \binits{E.E.}},
\bauthor{\bsnm{{Lundquist}}, \binits{L.L.}},
\bauthor{\bsnm{{Weber}}, \binits{M.}},
\bauthor{\bsnm{{Tarbell}}, \binits{T.D.}},
\bauthor{\bsnm{{Shine}}, \binits{R.A.}},
\bauthor{\bsnm{{S{\^o}ma}}, \binits{M.}},
\bauthor{\bsnm{{Tsuneta}}, \binits{S.}},
\bauthor{\bsnm{{Sakao}}, \binits{T.}},
\bauthor{\bsnm{{Minesugi}}, \binits{K.}}:
\byear{2007},
\batitle{{Hinode calibration for precise image co-alignment between SOT and XRT
  (2006 November-2007 April)}}.
\bjtitle{\pasj}
\bvolume{59},
\bfpage{S845}\,--\,\blpage{S852}.
\end{barticle}
\endbibitem

\bibitem[\protect\citeauthoryear{{Strong}
  \textit{et~al.}}{1992}]{Strongetal1992PASJ}
\begin{barticle}
\bauthor{\bsnm{{Strong}}, \binits{K.T.}},
\bauthor{\bsnm{{Harvey}}, \binits{K.}},
\bauthor{\bsnm{{Hirayama}}, \binits{T.}},
\bauthor{\bsnm{{Nitta}}, \binits{N.}},
\bauthor{\bsnm{{Shimizu}}, \binits{T.}},
\bauthor{\bsnm{{Tsuneta}}, \binits{S.}}:
\byear{1992},
\batitle{{Observations of the variability of coronal bright points by the Soft
  X-ray Telescope on YOHKOH}}.
\bjtitle{\pasj}
\bvolume{44},
\bfpage{L161}\,--\,\blpage{L166}.
\end{barticle}
\endbibitem

\bibitem[\protect\citeauthoryear{{Subramanian}, {Madjarska}, and
  {Doyle}}{2010}]{Subramanianetal2010A&A}
\begin{barticle}
\bauthor{\bsnm{{Subramanian}}, \binits{S.}},
\bauthor{\bsnm{{Madjarska}}, \binits{M.S.}},
\bauthor{\bsnm{{Doyle}}, \binits{J.G.}}:
\byear{2010},
\batitle{{Coronal hole boundaries evolution at small scales. II. XRT view. Can
  small-scale outflows at CHBs be a source of the slow solar wind?}}
\bjtitle{\aap}
\bvolume{516},
\bfpage{A50}.
doi:\doiurl{10.1051/0004-6361/200913624}.
\end{barticle}
\endbibitem

\bibitem[\protect\citeauthoryear{{Subramanian}
  \textit{et~al.}}{2012}]{Subramanianetal2012A&A}
\begin{barticle}
\bauthor{\bsnm{{Subramanian}}, \binits{S.}},
\bauthor{\bsnm{{Madjarska}}, \binits{M.S.}},
\bauthor{\bsnm{{Doyle}}, \binits{J.G.}},
\bauthor{\bsnm{{Bewsher}}, \binits{D.}}:
\byear{2012},
\batitle{{What is the true nature of blinkers?}}
\bjtitle{\aap}
\bvolume{538},
\bfpage{A50}.
doi:\doiurl{10.1051/0004-6361/201117877}.
\end{barticle}
\endbibitem

\bibitem[\protect\citeauthoryear{{Tian}
  \textit{et~al.}}{2008}]{Tianetal2008ApJ}
\begin{barticle}
\bauthor{\bsnm{{Tian}}, \binits{H.}},
\bauthor{\bsnm{{Curdt}}, \binits{W.}},
\bauthor{\bsnm{{Marsch}}, \binits{E.}},
\bauthor{\bsnm{{He}}, \binits{J.}}:
\byear{2008},
\batitle{{Cool and hot components of a coronal bright point}}.
\bjtitle{\apjl}
\bvolume{681},
\bfpage{L121}\,--\,\blpage{L124}.
doi:\doiurl{10.1086/590410}.
\end{barticle}
\endbibitem

\bibitem[\protect\citeauthoryear{{Ugarte-Urra}
  \textit{et~al.}}{2004}]{Ugarte-Urraetal2004A&A}
\begin{barticle}
\bauthor{\bsnm{{Ugarte-Urra}}, \binits{I.}},
\bauthor{\bsnm{{Doyle}}, \binits{J.G.}},
\bauthor{\bsnm{{Nakariakov}}, \binits{V.M.}},
\bauthor{\bsnm{{Foley}}, \binits{C.R.}}:
\byear{2004},
\batitle{{CDS wide slit time-series of EUV coronal bright points}}.
\bjtitle{\aap}
\bvolume{425},
\bfpage{1083}\,--\,\blpage{1095}.
doi:\doiurl{10.1051/0004-6361:20041069}.
\end{barticle}
\endbibitem

\bibitem[\protect\citeauthoryear{{Vlahos} and
  {Georgoulis}}{2004}]{Vlahosetal2004ApJ}
\begin{barticle}
\bauthor{\bsnm{{Vlahos}}, \binits{L.}},
\bauthor{\bsnm{{Georgoulis}}, \binits{M.K.}}:
\byear{2004},
\batitle{{On the self-similarity of unstable magnetic discontinuities in solar
  active regions}}.
\bjtitle{\apjl}
\bvolume{603},
\bfpage{L61}\,--\,\blpage{L64}.
doi:\doiurl{10.1086/383032}.
\end{barticle}
\endbibitem

\bibitem[\protect\citeauthoryear{{Webb}
  \textit{et~al.}}{1993}]{Webbetal1993SoPh}
\begin{barticle}
\bauthor{\bsnm{{Webb}}, \binits{D.F.}},
\bauthor{\bsnm{{Martin}}, \binits{S.F.}},
\bauthor{\bsnm{{Moses}}, \binits{D.}},
\bauthor{\bsnm{{Harvey}}, \binits{J.W.}}:
\byear{1993},
\batitle{{The correspondence between X-ray bright points and evolving magnetic
  features in the quiet sun}}.
\bjtitle{\solphys}
\bvolume{144},
\bfpage{15}\,--\,\blpage{35}.
doi:\doiurl{10.1007/BF00667979}.
\end{barticle}
\endbibitem

\bibitem[\protect\citeauthoryear{{Wilhelm}
  \textit{et~al.}}{1995}]{Wilhelmetal1995SoPh}
\begin{barticle}
\bauthor{\bsnm{{Wilhelm}}, \binits{K.}},
\bauthor{\bsnm{{Curdt}}, \binits{W.}},
\bauthor{\bsnm{{Marsch}}, \binits{E.}},
\bauthor{\bsnm{{Sch{\"u}hle}}, \binits{U.}},
\bauthor{\bsnm{{Lemaire}}, \binits{P.}},
\bauthor{\bsnm{{Gabriel}}, \binits{A.}},
\bauthor{\bsnm{{Vial}}, \binits{J.-C.}},
\bauthor{\bsnm{{Grewing}}, \binits{M.}},
\bauthor{\bsnm{{Huber}}, \binits{M.C.E.}},
\bauthor{\bsnm{{Jordan}}, \binits{S.D.}},
\bauthor{\bsnm{{Poland}}, \binits{A.I.}},
\bauthor{\bsnm{{Thomas}}, \binits{R.J.}},
\bauthor{\bsnm{{K{\"u}hne}}, \binits{M.}},
\bauthor{\bsnm{{Timothy}}, \binits{J.G.}},
\bauthor{\bsnm{{Hassler}}, \binits{D.M.}},
\bauthor{\bsnm{{Siegmund}}, \binits{O.H.W.}}:
\byear{1995},
\batitle{{SUMER - Solar Ultraviolet Measurements of Emitted Radiation}}.
\bjtitle{\solphys}
\bvolume{162},
\bfpage{189}\,--\,\blpage{231}.
doi:\doiurl{10.1007/BF00733430}.
\end{barticle}
\endbibitem

\bibitem[\protect\citeauthoryear{{Young} and
  {Gallagher}}{2008}]{2008SoPh..248..457Y}
\begin{barticle}
\bauthor{\bsnm{{Young}}, \binits{C.A.}},
\bauthor{\bsnm{{Gallagher}}, \binits{P.T.}}:
\byear{2008},
\batitle{{Multiscale edge detection in the corona}}.
\bjtitle{\solphys}
\bvolume{248},
\bfpage{457}\,--\,\blpage{469}.
doi:\doiurl{10.1007/s11207-008-9177-9}.
\end{barticle}
\endbibitem

\bibitem[\protect\citeauthoryear{{Young}
  \textit{et~al.}}{2007}]{Youngetal2007PASJ}
\begin{barticle}
\bauthor{\bsnm{{Young}}, \binits{P.R.}},
\bauthor{\bsnm{{Del Zanna}}, \binits{G.}},
\bauthor{\bsnm{{Mason}}, \binits{H.E.}},
\bauthor{\bsnm{{Doschek}}, \binits{G.A.}},
\bauthor{\bsnm{{Culhane}}, \binits{L.}},
\bauthor{\bsnm{{Hara}}, \binits{H.}}:
\byear{2007},
\batitle{{Solar transition region features observed with Hinode/EIS}}.
\bjtitle{\pasj}
\bvolume{59},
\bfpage{S727}\,--\,\blpage{S733}.
\end{barticle}
\endbibitem

\bibitem[\protect\citeauthoryear{{Zarro}}{2005}]{Zarro2005}
\begin{botherref}
\oauthor{\bsnm{{Zarro}}, \binits{D.M.}}:
2005,
\textit{{IDL Map Software for Analyzing Solar Images}},
http://orpheus.nascom.nasa.gov/zarro/idl/maps.html.
\end{botherref}
\endbibitem

\bibitem[\protect\citeauthoryear{{Zhang}, {Kundu}, and
  {White}}{2001}]{2001SoPh..198..347Z}
\begin{barticle}
\bauthor{\bsnm{{Zhang}}, \binits{J.}},
\bauthor{\bsnm{{Kundu}}, \binits{M.R.}},
\bauthor{\bsnm{{White}}, \binits{S.M.}}:
\byear{2001},
\batitle{{Spatial distribution and temporal evolution of coronal bright
  points}}.
\bjtitle{\solphys}
\bvolume{198},
\bfpage{347}\,--\,\blpage{365}.
doi:\doiurl{10.1023/A:1005222616375}.
\end{barticle}
\endbibitem

\bibitem[\protect\citeauthoryear{{Zhang}
  \textit{et~al.}}{2012}]{Zhangetal2012ApJ}
\begin{barticle}
\bauthor{\bsnm{{Zhang}}, \binits{Q.M.}},
\bauthor{\bsnm{{Chen}}, \binits{P.F.}},
\bauthor{\bsnm{{Guo}}, \binits{Y.}},
\bauthor{\bsnm{{Fang}}, \binits{C.}},
\bauthor{\bsnm{{Ding}}, \binits{M.D.}}:
\byear{2012},
\batitle{{Two types of magnetic reconnection in coronal bright points and the
  corresponding magnetic configuration}}.
\bjtitle{\apj}
\bvolume{746},
\bfpage{19}.
doi:\doiurl{10.1088/0004-637X/746/1/19}.
\end{barticle}
\endbibitem

\end{thebibliography}
\end{article}

\end{document}